\documentclass[showpacs,preprint,prb,aps]{revtex4-1}
\usepackage{amssymb}
\usepackage{epsfig}
\usepackage{subfigure}

\begin{document}
\title{A nanomechanical resonator coupled linearly via its momentum to a quantum point contact}

\author{Latchezar L. Benatov and Miles P. Blencowe}\affiliation{Department of Physics and Astronomy, Dartmouth College, Hanover, New Hampshire 03755, USA}

\date{\today}

\begin{abstract}
We use a Born-Markov approximated master equation approach to study the symmetrized-in-frequency current noise spectrum and the oscillator steady state of a nanoelectromechanical system where a nanoscale resonator is coupled linearly via its momentum to a quantum point contact (QPC).  Our current noise spectra exhibit clear signatures of the quantum correlations between the QPC current and the back-action force on the oscillator at a value of the relative tunneling phase ($\eta = -\pi/2$) where such correlations are expected to be maximized.  We also show that the steady state of the oscillator obeys a classical Fokker-Planck equation, but can experience thermomechanical noise squeezing in the presence of a momentum-coupled detector bath and a position-coupled environmental bath.  Besides, the full master equation clearly shows that half of the detector back-action is correlated with electron tunneling, indicating a departure from the model of the detector as an effective bath and suggesting that a future calculation valid at lower bias voltage, stronger tunneling and/or stronger coupling might reveal interesting quantum effects in the oscillator dynamics.
\end{abstract}


\maketitle

\section{\label{sec:introduction}Introduction}
Nanoelectromechanical systems (NEMS), in which a nano-to-micrometer scale mechanical resonator is coupled to an electronic device of similar dimensions, have received a great deal of theoretical and experimental attention in recent years, as these systems are a promising tool for gaining a deeper understanding of the quantum-to-classical transition in physics, in addition to their useful applications in ultrasensitive metrology.~\cite{blencowe05}  A wide variety of NEMS have already been realized experimentally, such as, for example, a doubly-clamped nanobeam coupled to a superconducting single-electron transistor (SSET),~\cite{naik06} a suspended carbon nanotube coupled to an embedded quantum dot~\cite{steele09} or single-electron transistor (SET),~\cite{lassagne09} a doubly-clamped beam coupled to an external SET,~\cite{knobel03} and a micromechanical cantilever coupled to a quantum point contact (QPC).~\cite{poggio08}  There have also been a number of theoretical studies of NEMS, in which the oscillator is coupled linearly via its position to a QPC~\cite{mozyrsky02,clerk04,bennett08,doiron09} or a SET.~\cite{armour04}  It has been shown both theoretically and experimentally that the effect of the electronic device (detector) on the oscillator is very similar to that of a thermal bath with a certain effective temperature and damping constant, even though the detector is in a far-from-equilibrium state.~\cite{blencowe05}  Besides, the oscillator can also have a strong effect on the detector, producing a Fano-like current noise spectrum.~\cite{rodrigues09} 

Most of the studies conducted so far have focused on a position-dependent linear coupling between the oscillator and the detector.  It is interesting to see how a $momentum$-dependent coupling changes the oscillator steady state and the detector current noise spectrum, and whether there is a nontrivial interplay between the effects of the momentum-coupled thermal bath associated with the detector and those of the position-coupled bath due to the environment of the oscillator.  Normally, when a position-coupled detector acts as a thermal bath with effective temperature $T_{\mathrm{det}}$ and damping constant $\gamma_{\mathrm{det}}$, in addition to the environmental bath temperature $T$ and damping $\gamma_0$, the oscillator is in a thermal state with effective damping $\gamma_{\mathrm{eff}} = \gamma_0 + \gamma_{\mathrm{det}}$ and temperature $T_{\mathrm{eff}} = (\gamma_0 T + \gamma_{\mathrm{det}} T_{\mathrm{det}})/\gamma_{\mathrm{eff}}$.~\cite{blencowe05}  In the absence of an environment, a momentum-coupled detector is equivalent to a position-coupled one under the canonical transformation, which interchanges the oscillator position and momentum coordinates.  However, the unavoidable presence of a position-coupled environmental bath breaks this symmetry, leading to potentially new and interesting physics. 
 
\begin{figure}[htbp]
	\centering
		\includegraphics[width=0.75\textwidth]{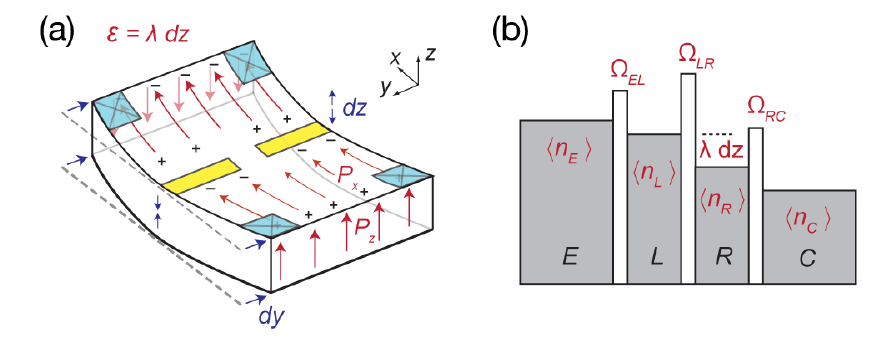}
	\caption{(a) Displacement and associated polarization of the GaAs crystal (Fig. 1b in Ref.~[\onlinecite{stettenheim10}]).  (b) Schematic diagram of the different electronic reservoirs and tunnel barriers in the system (Fig. 1c in Ref.~[\onlinecite{stettenheim10}]).  The symbols are defined in the text.}
	\label{Fig1}
\end{figure}

One example of a NEMS which, after an appropriate transformation (see Sec.~\ref{sec:mastereq} below), can be described by a momentum-coupled effective Hamiltonian, was studied experimentally by Stettenheim et al.~\cite{stettenheim10}  Their experiment involved a nanomechanical GaAs oscillator coupled piezoelectrically to a radio-frequency QPC embedded in it. Measurements of the current noise through the QPC detector showed that the quantum statistical fluctuations of tunneling electrons could affect the macroscopic dynamics of the host crystal.  Fig.~\ref{Fig1}(a) (reproduced from Ref.~[\onlinecite{stettenheim10}]) shows the GaAs crystal containing a two-dimensional electron gas (2DEG).  A displacement $dy$ of the front and back faces of the crystal leads to a compression $dz$ at the midpoint of the left face and a corresponding  expansion $dz$ at the midpoint of the right face, as shown.  The resulting strain $S_{yz} = 2dz/w$, where $w$ is the width of the crystal, produces, through the piezoelectric coupling constant $e_{x4}$, a bulk polarization $P_x = e_{x4} S_{yz}$, which is assumed to be in the direction of transport through the QPC.  The 2DEG electrons will try to screen the polarization charge, but under the gates and in the QPC, where the 2DEG is depleted, there will be a net electric field and a corresponding potential difference $d \epsilon = \lambda dz$ between the left ($L$) and right ($R$) reservoirs, leading to a current $I$ through the QPC.  One of the normal vibrational modes of GaAs has a polarization field as in Fig.~\ref{Fig1}(a), and thus the QPC current $I$ can provide information on the displacement $dz$ of the crystal as it oscillates in this mode.  On the other hand, the unavoidable shot noise due to partitioning of electron-hole pairs at the QPC leads to charge fluctuations $dn$ in reservoirs $L$ and $R$, and a corresponding back-action force $dF = \eta dn$ on the oscillator via the piezoelectric effect, completing a feedback loop between the mechanical and electronic degrees of freedom.  Thus one expects both the mechanical motion of the resonator and the current noise through the QPC detector to be peaked at the oscillator frequency.  An interesting result in the experiment, which we set out to investigate theoretically in the present study, is that the current noise spectrum of the QPC displays super-Poissonian values close to the oscillator frequency yet sub-Poissonian values away from it, indicating bunching and anti-bunching of electron tunneling events due to the coupling to the oscillator.  One important caveat to keep in mind when comparing theory and experiment, however, is that the experiment was performed in the strong tunneling regime, where the QPC conductance $G_{\mathrm{QPC}} \approx 0.5 G_0$ and $G_0 = 2e^2/h$ is the conductance quantum, whereas our theoretical calculation is based on the assumption of weak tunneling.  The case of strong tunneling will be considered in a future publication. 

The present paper is structured as follows: In Sec.~\ref{sec:mastereq}, we perform a polaron-like transformation on the Hamiltonian of the system described above, leading to an oscillator momentum-dependent tunneling amplitude across the QPC, and derive a Born-Markov approximated master equation for the reduced oscillator density matrix.  In Sec.~\ref{sec:currentnoise}, we solve the master equation to obtain the average current and symmetrized-in-frequency current noise spectrum through the QPC.  The noise spectra thus obtained are investigated in Sec.~\ref{sec:results} for a wide range of system parameters.  In Sec.~\ref{sec:steadystate}, we use the Wigner representation of quantum mechanics to study the steady state of the oscillator.  We present our conclusions in Sec.~\ref{sec:conclusion}.  Details of the polaron transformation, the derivation of the Born-Markov master equation, and its solution to obtain the current noise spectrum are presented in Appendices~\ref{sec:appendixa},~\ref{sec:appendixb} and~\ref{sec:appendixc}, respectively.

\section{\label{sec:mastereq}Derivation of the Born-Markov Master Equation for the Fourier-Transformed Reduced Density Matrix of the Oscillator}

As explained in the introduction, the piezoelectric coupling between the 2DEG reservoirs and the flexing GaAs crystal can be modeled by a linear dependence of the single electron energy levels in the $L$ and $R$ reservoirs on the crystal displacement (Fig.~\ref{Fig1}(b)).  The energy levels of the emitter ($E$) and collector ($C$) reservoirs, which represent the leads connected via tunnel barriers to the $L$ and $R$ reservoirs, respectively, are assumed to be fixed.  Thus one starts with the Hamiltonian $H = H_{\mathrm{sys}} + H_{\mathrm{bath}} + H_{\mathrm{int}}$, where the system, interaction, and bath Hamiltonians are, respectively:
\begin{eqnarray}
H_{\mathrm{sys}} & = & \hbar \omega_m a^{\dagger} a + \sum_L (\epsilon_L - \lambda x/2) b_L^{\dagger} b_L + \sum_R (\epsilon_R + \lambda x/2) b_R^{\dagger} b_R, \nonumber \\
H_{\mathrm{bath}} & = & \sum_E \epsilon_E b_E^{\dagger} b_E + \sum_C \epsilon_C b_C^{\dagger} b_C, \nonumber \\
H_{\mathrm{int}} & = & \sum_{E, L} \hbar \Omega_{EL} b_E^{\dagger} b_L + \sum_{C, R} \hbar \Omega_{CR} b_C^{\dagger} b_R Y^{\dagger} + \sum_{L,R} \hbar \Omega_{LR} b_L^{\dagger} b_R + \mathrm{H.c.}
\label{orig_Ham}
\end{eqnarray}
Here the operators $b_i$ ($b_i^{\dagger}$) denote the annihilation (creation) operators for the energy levels $\epsilon_i$ of a given reservoir  $i = E, L, R, C$, the $Y$ operator counts electrons traversing the $RC$ reservoir barrier, the $\Omega_{ij}$ are the tunneling amplitudes between the various adjacent reservoirs, the parameter $\lambda$ describes the piezoelectric coupling between the (bosonic) crystal vibrational mode $x = x_{\mathrm{zp}} (a + a^{\dagger})$ with frequency $\omega_m$ and the $L$, $R$ reservoir electrons, and $x_{\mathrm{zp}}$ is the vibrational amplitude zero-point uncertainty.  It is important to note that we use $x$, and not $z$ as in Ref.~[\onlinecite{stettenheim10}] and Fig.~\ref{Fig1}, as the direction of motion of the oscillator throughout the rest of this paper.

The coupling described above is rather unusual - in most electromechanical systems studied so far, the oscillator position affects either the charge state of a single island,~\cite{naik06,steele09,lassagne09,knobel03,armour04} or the tunnel-barrier potential,~\cite{poggio08,mozyrsky02,clerk04,bennett08,doiron09} rather than the reservoirs to the left and right of the tunnel barrier.  To obtain the current noise spectrum, one could derive a Born-Markov approximated master equation directly from the above Hamiltonian, tracing over the bath degrees of freedom comprising the $E$ and $C$ reservoir electrons.  However, this approach can be quite involved due to the large number of coupled second order moment equations that one needs to solve.  An easier method is to perform a polaron-like transformation on Eq.~(\ref{orig_Ham}), and derive a much simpler effective Hamiltonian for our system, in which the coupling maps effectively onto a $momentum$-dependent tunnel barrier potential.  We replace $H \rightarrow U H U^{\dagger}$, where the unitary operator is
\begin{equation} 
U = \exp \Bigg[ - \frac{\lambda x_{\mathrm{zp}}}{2 \hbar \omega_m} \Bigg( \sum_L b_L^{\dagger} b_L - \sum_R b_R^{\dagger} b_R \Bigg) (a^{\dagger} - a) \Bigg].
\end{equation}
Expanding to first order in the oscillator displacement and neglecting the quartic terms in the $b_L$ and $b_R$ operators as well as the momentum dependence in the $E$ and $C$ contact resistance barrier terms, we arrive at the much simpler Hamiltonian $H = H_{\mathrm{osc}} + H_{\mathrm{bath}} + H_{\mathrm{int}}$, where
\begin{eqnarray}
H_{\mathrm{osc}} & = & \hbar \omega_m a^{\dagger} a, \nonumber \\
H_{\mathrm{bath}} & = & \sum_L \epsilon_L b_L^{\dagger} b_L + \sum_R \epsilon_R b_R^{\dagger} b_R, \nonumber \\
H_{\mathrm{int}} & = & \Bigg[ 1 - \frac{\lambda x_{\mathrm{zp}}}{\hbar \omega_m} (a^{\dagger} - a) \Bigg] \sum_{L,R} \hbar \Omega_{LR} b_L^{\dagger} b_R Y + \mathrm{H.c.},
\end{eqnarray}
and $Y$ now counts electrons tunneling through the $LR$ reservoir barrier.  This calculation is described in detail in Appendix~\ref{sec:appendixa}.  In the above derivation, we have assumed weak coupling between the oscillator and QPC, $\tilde{\lambda} \ll 1$, where the dimensionless coupling parameter is
\begin{equation}
\tilde{\lambda} = \frac{\lambda}{\sqrt{2 \hbar m \omega_m^3}}
\label{dimlesscoupling}
\end{equation}
and $m$ is the oscillator mass, as well as weak tunneling, $t_0 \ll 1$, where the bare tunneling amplitude $t_0$ is defined below in Eq.~(\ref{tunnelingpar}).  Together with the assumption of high bias voltage across the QPC, $eV/\hbar \omega_m \gg 1$, needed to make the Born-Markov approximation described below, these are the three main conditions of validity of our calculation.  As already mentioned, the experiment of Ref.~[\onlinecite{stettenheim10}] is in the regime of $strong$ tunneling, where the polaronic and Born-Markov approximations are no longer valid and scattering matrix methods can be used instead~\cite{bennett10}  - an approach we intend to investigate in future work. 

It is convenient to express $H_{\mathrm{int}}$ in terms of the oscillator momentum, $\hat{p} = (i \hbar / 2 x_{\mathrm{zp}}) (a^{\dagger} - a)$:
\begin{equation}
H_{\mathrm{int}} = \hat{T}(p) \sum_{L,R} b_R^{\dagger} b_L + \mathrm{H.c.},
\end{equation}
where 
\begin{equation}
\hat{T}(p) = \hbar \Omega_{LR}^{\ast} \bigg( 1 - i \frac{\lambda}{\hbar m \omega_m^2} \hat{p} \bigg) Y^{\dagger},
\end{equation}
and we have assumed that $\Omega_{LR}$ is level-independent.  From this point, we follow the approach of Doiron,~\cite{doiron09} since our Hamiltonian has exactly the same form as his, except that his tunneling amplitude $\hat{T}(x)$ is position-dependent.  We can write 
\begin{equation}
\hat{T}(p) = \frac{1}{2 \pi \Lambda} (t_0 + e^{i \eta} t_1 \hat{p}) Y^{\dagger},
\end{equation}
where $\Lambda$ is the constant density of states in the reservoirs, and
\begin{eqnarray}
t_0 & = & 2 \pi \Lambda \hbar |\Omega_{LR}|, \nonumber \\
t_1 & = & \frac{2 \pi \Lambda |\Omega_{LR}| \lambda}{m \omega_m^2}, \nonumber \\
\eta & = & - \frac{\pi}{2}.
\label{tunnelingpar}
\end{eqnarray}
We have taken the absolute value of $\Omega_{LR}^{\ast}$ since the overall phase of $t_0$ and $t_1$ is unimportant and only the relative phase difference $\eta$ matters physically.  In Ref.~[\onlinecite{clerk04}] and most other studies so far, only the case of $\eta = 0$ was investigated for the position-coupled system, implying a zero average back-action force on the oscillator.  In our case, $\eta = -\pi/2$ implies that the average back-action force is non-zero (cf.~Eq.~(\ref{backaction}) below; under the canonical transformation, which interchanges the oscillator position and momentum coordinates,  $\overline{F}_0(\eta)$ indeed becomes the average back-action force, cf.~the first line of Eq.~(\ref{eq:master2}) below).  The case of non-zero $\eta$ is considered in Ref.~[\onlinecite{doiron08}], where two tunnel junctions, one of which is linearly coupled to an oscillator via its position, are arranged in an Aharonov-Bohm-type setup, and the magnetic flux through the loop can be used to tune the phase $\eta$ between the oscillator-independent and oscillator-dependent total tunneling amplitudes.  It is shown that when $\eta = 0$ mod $\pi$,  the current noise spectrum of the detector is proportional to the position spectrum of the oscillator as in Ref.~[\onlinecite{clerk04}], but when $\eta = \pi/2$ mod $\pi$, the noise spectrum is proportional to the $momentum$ spectrum of the oscillator.  On the other hand, in Ref.~[\onlinecite{walter11}] it is demonstrated that for a $non$-$stationary$ oscillator coupled to a single QPC via its position, the current noise spectrum of the detector is complex-valued and contains information about both the oscillator position and the oscillator momentum, even when $\eta = 0$.  

Assuming that the oscillator-bath coupling is weak and the bath correlations decay much faster than the characteristic timescale of the oscillator, we can use a Born-Markov approximation technique to derive a master equation for the Fourier-transformed reduced oscillator density matrix
\begin{equation}
\rho(\chi;t) = \sum_N e^{i \chi N} \rho(N;t),
\label{NchiFT}
\end{equation}
where $\rho(N;t) = \langle N | \rho_{\mathrm{osc}}(t) | N \rangle$ is the $N$-resolved oscillator density matrix and $N$ is the number of electrons that have tunneled from the left into the right lead at time $t$.  The details of this calculation are presented in Appendix~\ref{sec:appendixb}, and the final result is
\begin{eqnarray}
\frac{d}{dt} \rho(\chi;t) & = & \frac{1}{i \hbar} [H_{\mathrm{osc}} - \overline{F}_{\mathrm{0}}(-\frac{\pi}{2}) \hat{p}, \rho(\chi;t)] - \sum_{\sigma = \pm 1} \frac{D_{\sigma}}{\hbar^2} [\hat{p}, [\hat{p}, \rho(\chi;t)]] \nonumber \\ 
& + & \frac{i}{\hbar} m^2 \omega_m^2 \sum_{\sigma = \pm 1} \tilde{\gamma}_{\sigma} [\hat{p}, \{\hat{x}, \rho(\chi;t) \} ] \nonumber \\
& - & \frac{D_0}{\hbar^2} [\hat{x}, [\hat{x}, \rho(\chi;t)]] - \frac{i}{\hbar} \tilde{\gamma}_0 [\hat{x}, \{\hat{p}, \rho(\chi;t) \} ] \nonumber \\
& + & \sum_{\sigma = \pm 1} \Gamma_{\sigma}(0) \big(e^{i \chi \sigma} - 1 \big) \bigg( \rho(\chi;t) - \frac{\sigma i t_1}{2 t_0} [\hat{p}, \rho(\chi;t)] \bigg) \nonumber \\
& - & \sum_{\sigma = \pm 1} \frac{\sigma i D_{\sigma}}{\hbar^2} \bigg( \frac{e^{i \chi \sigma} - 1}{t_1^2} \bigg) \big(t_0 t_1 [\hat{p}, \rho(\chi;t)] \big) \nonumber \\
& + & \sum_{\sigma = \pm 1} \frac{2 D_{\sigma}}{\hbar^2} \bigg( \frac{e^{i \chi \sigma} - 1}{t_1^2} \bigg) \big(t_1^2 \big(\hat{p} \rho(\chi;t) \hat{p} \big) \big) \nonumber \\
& - & \sum_{\sigma = \pm 1} \frac{\sigma m^2 \omega_m^2 \tilde{\gamma}_{\sigma}}{\hbar} \bigg( \frac{e^{i \chi \sigma} - 1}{t_1^2} \bigg) \big(t_0 t_1 \{ \hat{x}, \rho(\chi;t) \} \big) \nonumber \\
& + & \sum_{\sigma = \pm 1} \frac{i m^2 \omega_m^2 \tilde{\gamma}_{\sigma}}{\hbar} \bigg( \frac{e^{i \chi \sigma} - 1}{t_1^2} \bigg) \big( t_1^2 \big( \hat{p} \rho(\chi;t) \hat{x} - \hat{x} \rho(\chi;t) \hat{p} \big) \big).
\label{eq:master2}
\end{eqnarray}
We have defined
\begin{eqnarray}
\tilde{\gamma}_{\sigma} & = & \frac{\hbar}{4 m \omega_m} \bigg( \frac{t_1}{t_0} \bigg)^2 [\Gamma_{\sigma}(\hbar \omega_m) - \Gamma_{\sigma}(-\hbar \omega_m)], \\
D_{\sigma} & = & \frac{\hbar^2}{4} \bigg( \frac{t_1}{t_0} \bigg)^2 [\Gamma_{\sigma}(\hbar \omega_m) + \Gamma_{\sigma}(-\hbar \omega_m)], \\
\overline{F}_0(\eta) & = & \hbar \sin \eta \bigg( \frac{t_1}{t_0} \bigg) \sum_{\sigma} \sigma \Gamma_{\sigma}(0),
\label{backaction}
\end{eqnarray}
where $\sigma = \pm 1$ and the forward (left to right) and backward (right to left) tunneling rates are, respectively, 
\begin{equation}
h \Gamma_{+}(E) = \int_0^\infty d \epsilon |t_0|^2 f(\epsilon - \mu_L) [1 - f(\epsilon - \mu_R + E)],
\label{forwardrate}
\end{equation}
\begin{equation}
h \Gamma_{-}(E) = \int_0^\infty d \epsilon |t_0|^2 f(\epsilon - \mu_R) [1 - f(\epsilon - \mu_L + E)], 
\label{backwardrate}
\end{equation}
where $\mu_i$ and $f_i$ are the chemical potential and Fermi function of reservoir $i$.  We have also modeled the environment of the oscillator as a thermal bath by including external diffusion and damping terms (the first and second terms on the third line of Eq.~(\ref{eq:master2}), respectively), where
\begin{eqnarray}
D_0 & = & m \tilde{\gamma}_0 \hbar \omega_m \coth \bigg( \frac{\hbar \omega_m}{2 k_B T} \bigg) \nonumber \\
& \approx & 2 m \tilde{\gamma}_0 k_B T \qquad (\textrm{when} \quad k_B T \gg \hbar \omega_m), 
\label{ext_diffusion}
\end{eqnarray}
$\tilde{\gamma}_0$ is the external oscillator damping constant (defined in such a way that $-2 \tilde{\gamma}_0 p$ is the classical external damping force), and $T$ is the temperature of the environment. 

If the external damping and diffusion terms were not present, our system would be identical under the canonical transformation ($\hat{p} \leftrightarrow m \omega_m \hat{x}$) to the position-coupled resonator-QPC systems studied by other groups.~\cite{poggio08,mozyrsky02,clerk04,bennett08,doiron09}, but with non-zero $\eta$.  However, the presence of such terms destroys this correspondence and creates a fundamentally new situation, in which there is a potentially interesting interplay between the effects of the position-coupled environment and those of the momentum-coupled detector on the oscillator.

\section{\label{sec:currentnoise}Solving the Master Equation to Find the Average Current and Symmetrized-in-Frequency Current Noise Spectrum}
We assume a large forward bias (so that we can drop the $\sigma = -1$ terms in Eq.~(\ref{eq:master2})) and zero temperature in the leads so that we can set $f(\epsilon - \mu_{L,R}) = \Theta(\mu_{L,R} - \epsilon)$ in the definitions of $\Gamma_{\pm}(E)$ and compute the resulting simple integrals in Eqs.~(\ref{forwardrate})-(\ref{backwardrate}).  The average current and current noise can be computed from the moments of $N$ using the formula
\begin{equation}
\langle N^n(t) \rangle = i^{-n} \mathrm{Tr} \bigg( \frac{d^n}{d \chi^n} \rho(\chi;t) \bigg)_{\chi = 0}
\end{equation}
and taking the $\chi$-derivative and trace of Eq.~(\ref{eq:master2}) (cf. Eq.~(\ref{NchiFT})).  After a straightforward calculation we find for the average current
\begin{equation}
\langle I \rangle = e \frac{d}{dt} \langle N \rangle = \frac{e^2 V}{h} \big(t_0^2 + t_1^2 \langle p^2 \rangle \big) - \frac{2 e m^2 \omega_m^2}{\hbar} \frac{t_0}{t_1} \tilde{\gamma}_{+} \langle x \rangle - e m^2 \omega_m^2 \tilde{\gamma}_{+},
\end{equation}
where $\tilde{\gamma}_{+} = \hbar^2 t_1^2/2mh$ and $eV = \mu_L - \mu_R$ is the QPC bias voltage.  To find the symmetrized-in-frequency current noise spectrum, 
\begin{equation}
\bar{S}_I(\omega) = \frac{1}{2} \int_{-\infty}^{\infty} \langle \{ \delta I(\tau), \delta I(0) \} \rangle e^{i \omega \tau} d \tau,
\end{equation}
where $\delta I = I - \langle I \rangle$, we use the MacDonald formula,~\cite{macdonald49}
\begin{equation}
\bar{S}_I(\omega) = 2 e^2 \omega \int_0^\infty dt \sin(\omega t) \frac{d}{dt} \langle \langle N^2(t) \rangle \rangle.
\end{equation}
The time derivative of the variance of $N$ can be computed from the time derivatives of the moments using the expression
\begin{equation}
\frac{d}{dt} \langle \langle N^2(t) \rangle \rangle = \frac{d}{dt} \langle N^2(t) \rangle - 2 \langle N(t) \rangle \frac{d}{dt} \langle N(t) \rangle.
\end{equation}
One obtains the following result:
\begin{equation}
\bar{S}_I = 2 e \langle I \rangle + \Delta \bar{S}_I,
\end{equation}
where the first term is the Poissonian (oscillator-independent) part of the noise, and the oscillator-dependent part is given by the integral
\begin{equation}
\Delta \bar{S}_I = 2 e^2 \omega \int_0^\infty dt \sin(\omega t) \bigg( - \frac{4 m^2 \omega_m^2 \tilde{\gamma}_{+}}{\hbar} \frac{t_0}{t_1} \langle \langle x N \rangle \rangle + \frac{2 e V}{h} t_1^2 \langle \langle p^2 N \rangle \rangle \bigg).
\label{eq:deltaSI}
\end{equation}
It is possible to solve analytically for the time dependence of the cumulants $\langle \langle x N \rangle \rangle$ and $\langle \langle p^2 N \rangle \rangle$, and integrate them to obtain an algebraic expression for $\Delta \bar{S}_I$.  (Note: The double angular brackets used throughout this article denote second cumulants, i.e. covariances, between products of powers of the oscillator coordinates (e.g. $x$, $p^2$ or $xp$) and $N$, not higher order cumulants.)  The full calculation and results are presented in Appendix~\ref{sec:appendixc}. 

\section{\label{sec:results}Results}
To plot the current noise spectrum, it is useful to put the equations in Appendix~\ref{sec:appendixc} in dimensionless form.  We define
\begin{eqnarray}
\tilde{x} & = & \frac{x}{x_{\mathrm{zp}}}, \nonumber \\
\tilde{p} & = & \frac{p}{p_{\mathrm{zp}}}, \nonumber \\
\tau & = & \omega_m t, \nonumber \\
\tilde{\omega} & = & \frac{\omega}{\omega_m}
\end{eqnarray}
to be the dimensionless oscillator position, oscillator momentum, time and oscillator frequency, respectively, where
\begin{eqnarray}
x_{\mathrm{zp}} & = & \sqrt{\frac{\hbar}{2 m \omega_m}}, \nonumber \\
p_{\mathrm{zp}} & = & \sqrt{\frac{\hbar m \omega_m}{2}}
\end{eqnarray}
are the oscillator position and momentum zero-point uncertainties.  Our system is then governed by five dimensionless parameters:
\begin{eqnarray}
\tilde{V} & = & \frac{eV}{\hbar \omega_m}, \nonumber \\
\tilde{\lambda} & = & \frac{\lambda}{\sqrt{2 \hbar m \omega_m^3}}, \nonumber \\
\tilde{\Gamma}_0 & = & \frac{\tilde{\gamma}_0}{\omega_m}, \nonumber \\
\tilde{T} & = & \frac{k_B T}{\hbar \omega_m}
\label{dimlesspar}
\end{eqnarray}
and $t_0$, i.e. the dimensionless bias voltage, coupling, external damping, external temperature, and bare tunneling amplitude. 

\begin{figure}[htbp]
\begin{center}
\includegraphics[width=0.75\textwidth]{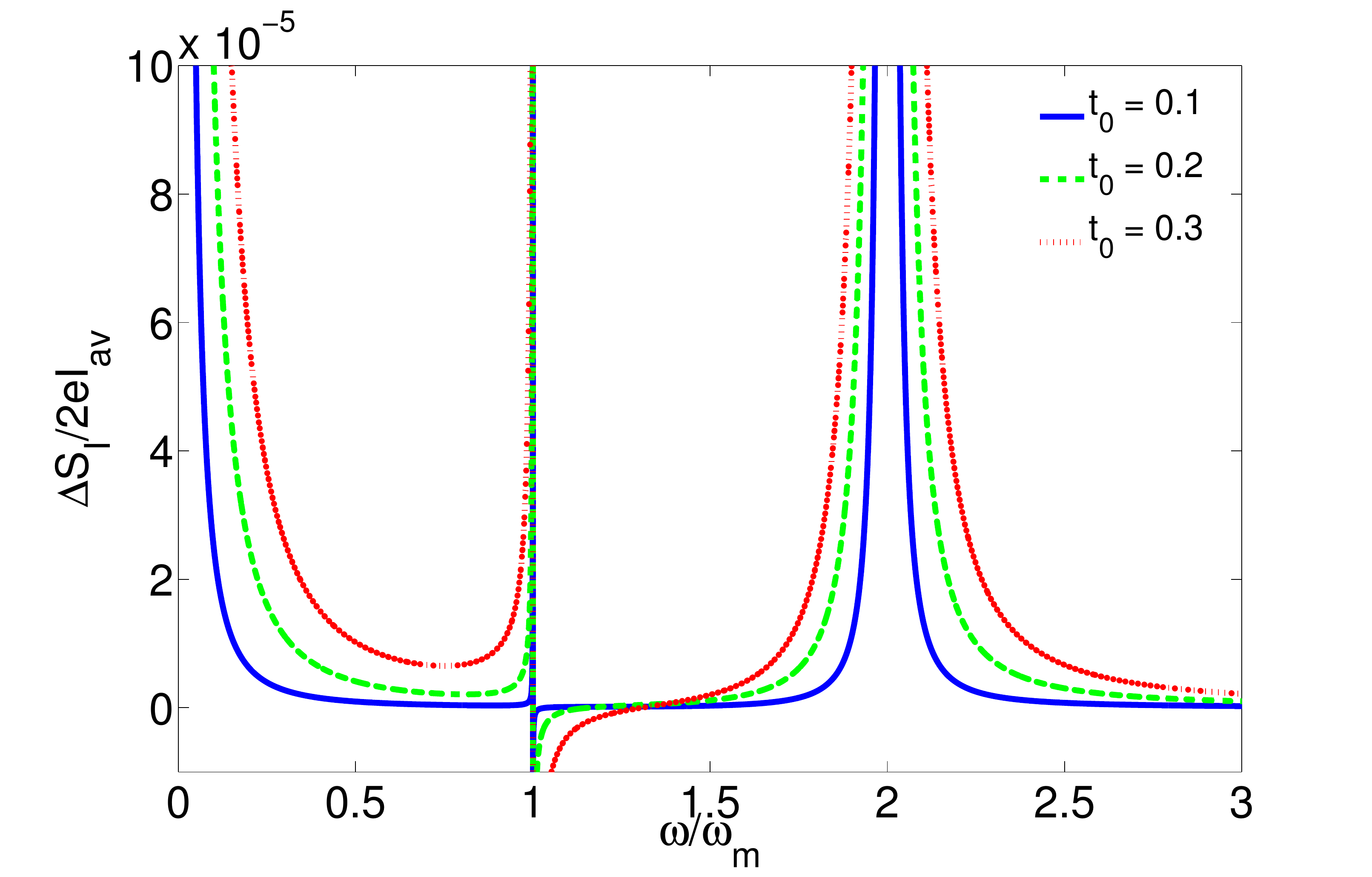}
\end{center}
\caption{Dimensionless non-Poissonian current noise spectrum for $t_0 = 0.1$, $0.2$ and $0.3$.  The remaining parameters are $\tilde{V} = 2 \times 10^4$, $\tilde{\lambda} = 1 \times 10^{-3}$, $\tilde{\Gamma}_0 = 5 \times 10^{-6}$, and $\tilde{T} = 1 \times 10^4$.}
\label{tunneling}
\end{figure}

\begin{figure}[htbp]
\begin{center}
\includegraphics[width=0.75\textwidth]{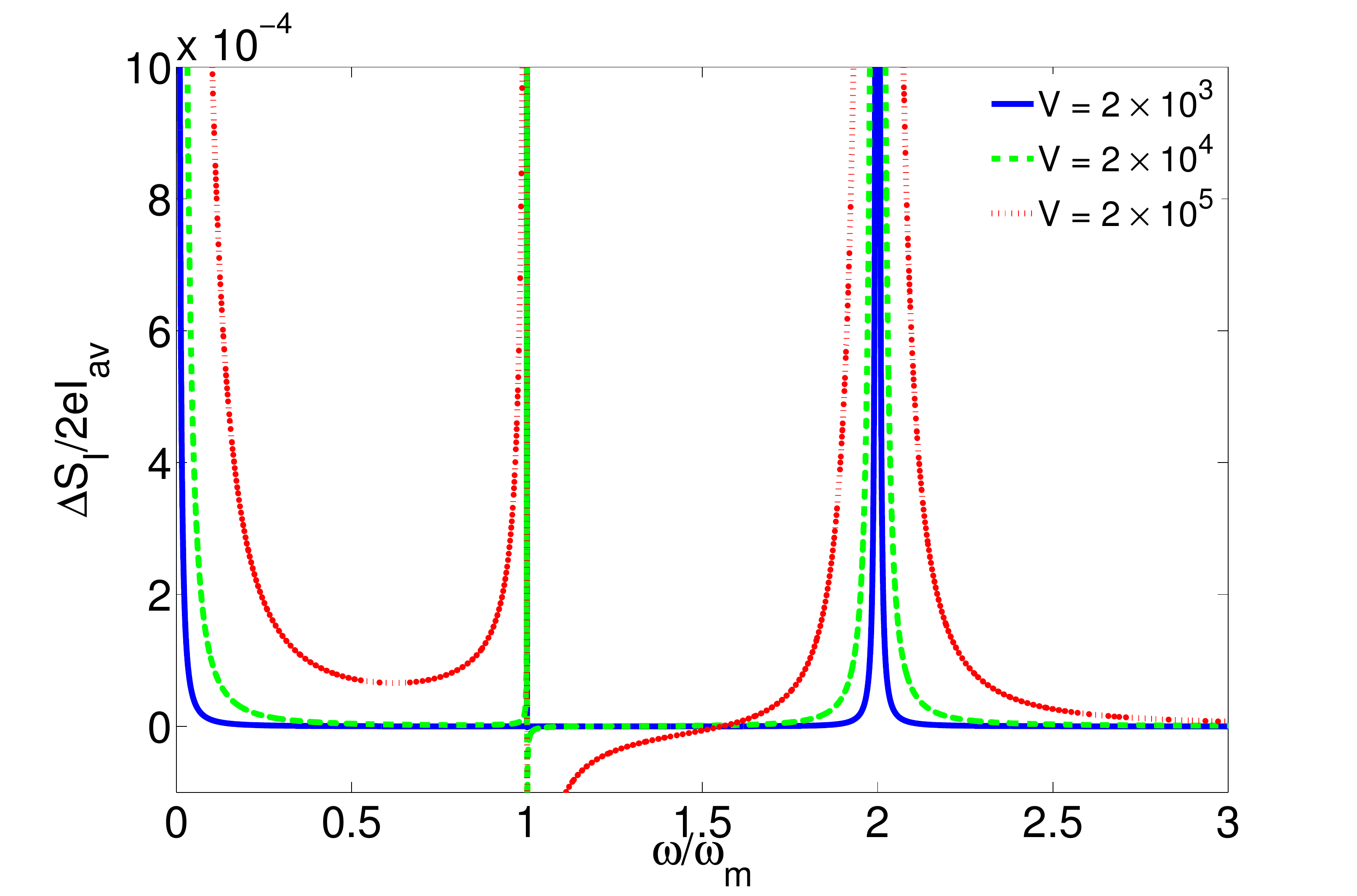}
\end{center}
\caption{Dimensionless non-Poissonian current noise spectrum for $\tilde{V} = 2 \times 10^3$, $2 \times 10^4$ and $2 \times 10^5$.  The remaining parameters are $t_0 = 0.2$, $\tilde{\lambda} = 1 \times 10^{-3}$, $\tilde{\Gamma}_0 = 5 \times 10^{-6}$, and $\tilde{T} = 1 \times 10^4$.}
\label{voltage}
\end{figure}

\begin{figure}[htbp]
\begin{center}
\includegraphics[width=0.75\textwidth]{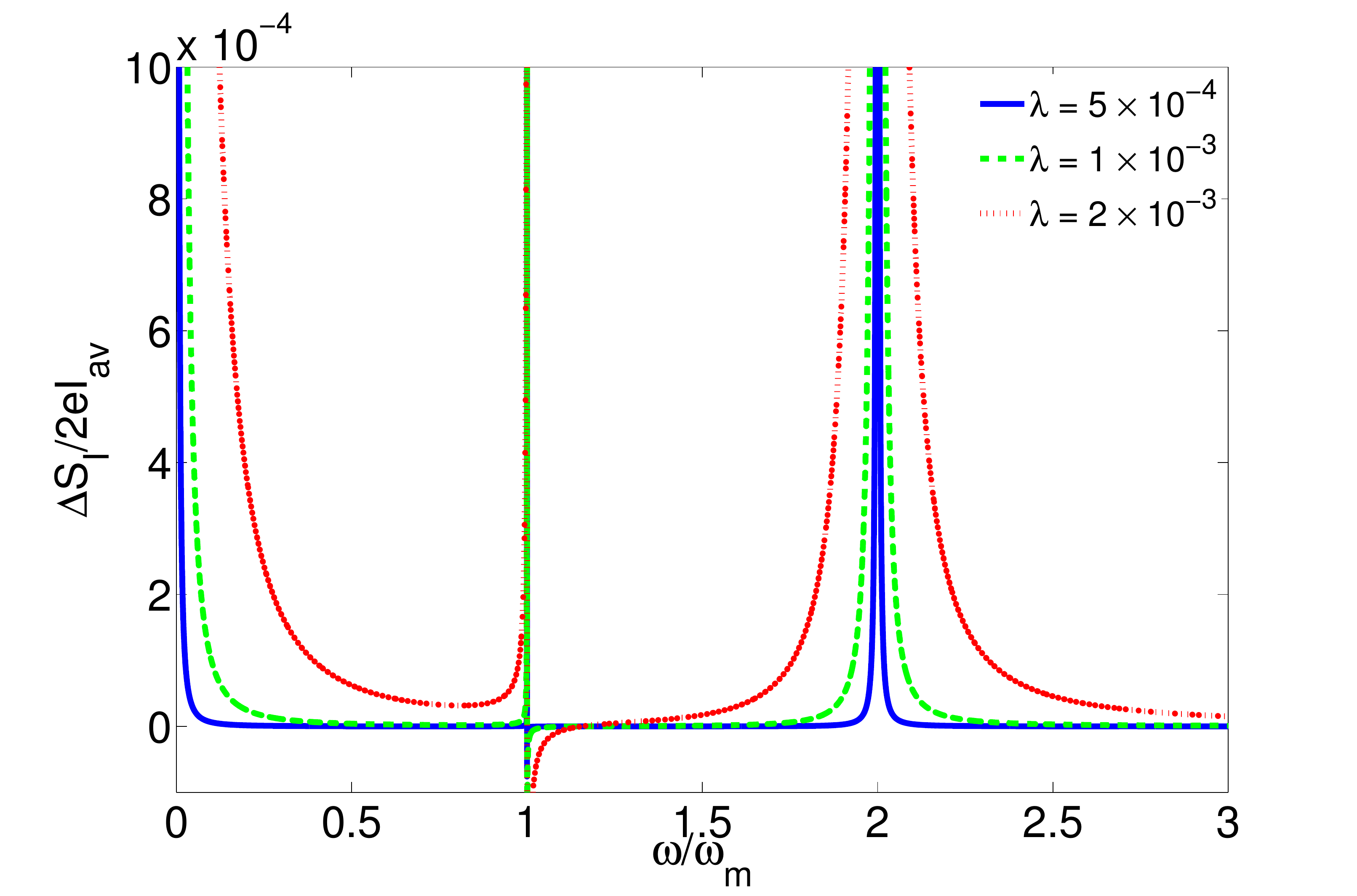}
\end{center}
\caption{Dimensionless non-Poissonian current noise spectrum for $\tilde{\lambda} = 5 \times 10^{-4}$, $1 \times 10^{-3}$ and $2 \times 10^{-3}$.  The remaining parameters are $t_0 = 0.2$, $\tilde{V} = 2 \times 10^4$, $\tilde{\Gamma}_0 = 5 \times 10^{-6}$, and $\tilde{T} = 1 \times 10^4$.}
\label{coupling}
\end{figure}

\begin{figure}[htbp]
\begin{center}
\includegraphics[width=0.75\textwidth]{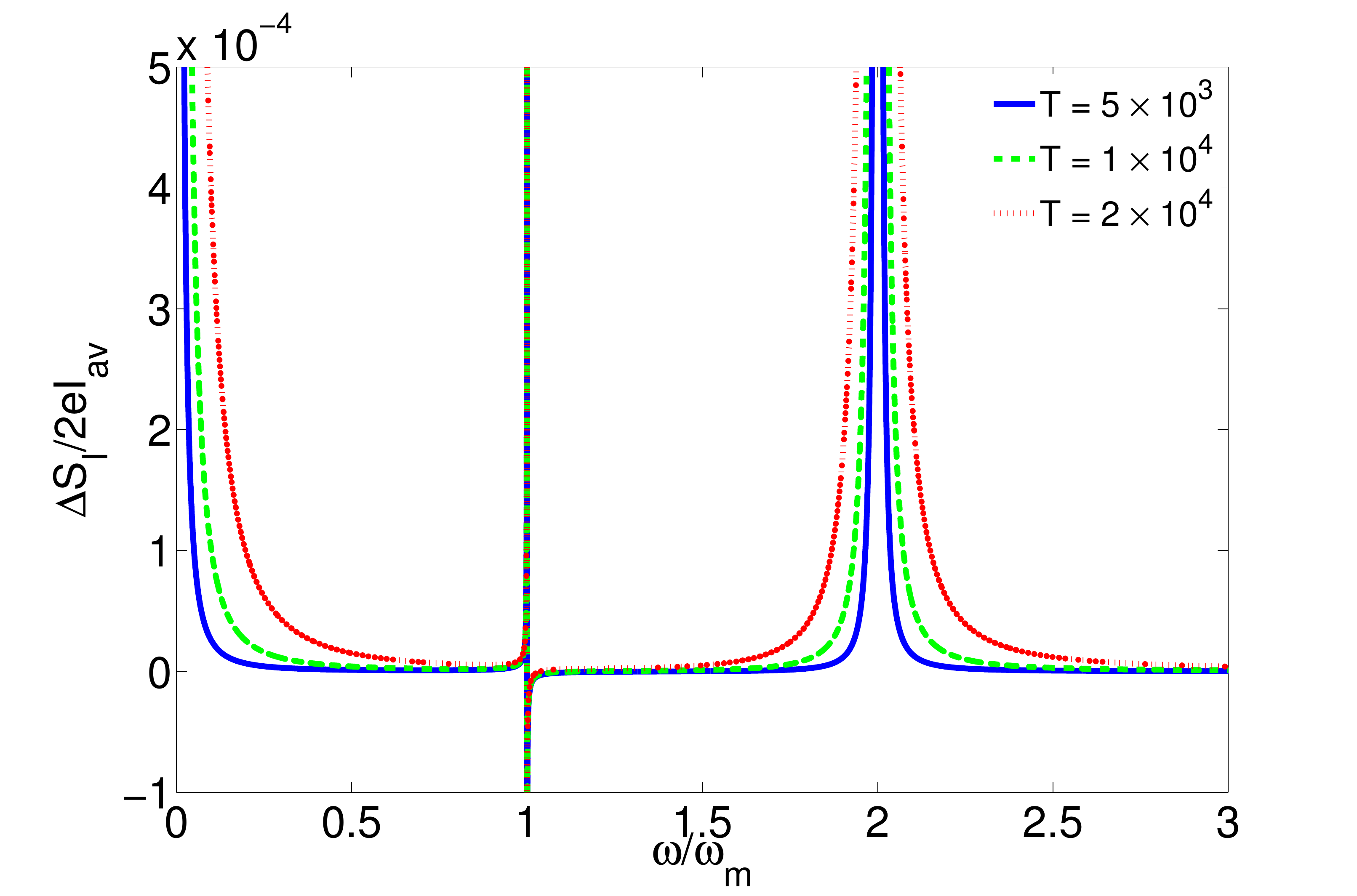}
\end{center}
\caption{Dimensionless non-Poissonian current noise spectrum for $\tilde{T} = 5 \times 10^3$, $1 \times 10^4$ and $2 \times 10^4$.  The remaining parameters are $t_0 = 0.2$, $\tilde{V} = 2 \times 10^4$, $\tilde{\lambda} = 1 \times 10^{-3}$, and $\tilde{\Gamma}_0 = 5 \times 10^{-6}$.}
\label{temp}
\end{figure}

\begin{figure}[htbp]
\begin{center}
\includegraphics[width= 0.75\textwidth]{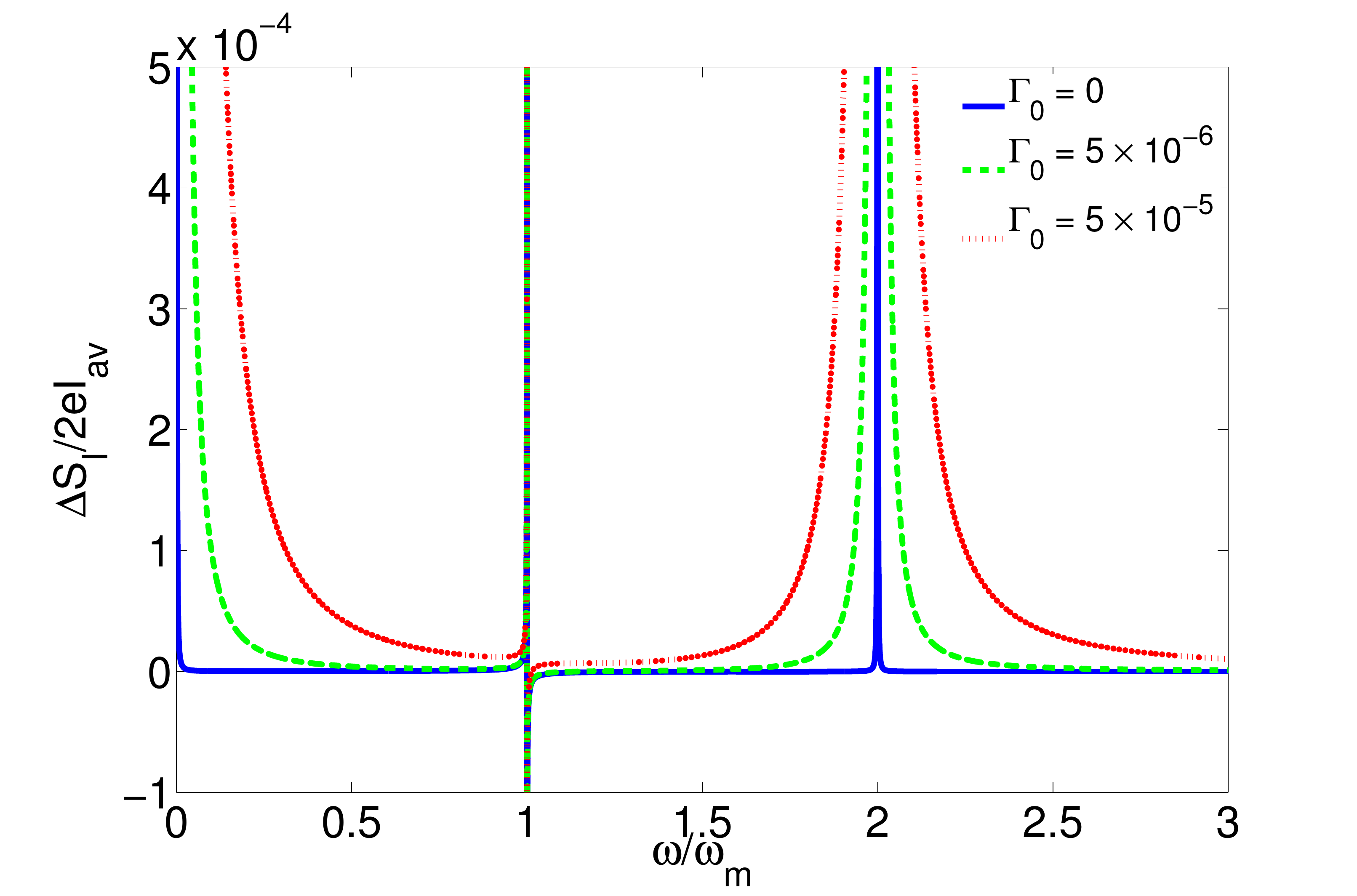}
\end{center}
\caption{Dimensionless non-Poissonian current noise spectrum for $\tilde{\Gamma}_0 = 0$, $5 \times 10^{-6}$ and $5 \times 10^{-5}$.  The remaining parameters are $t_0 = 0.2$, $\tilde{V} = 2 \times 10^4$, $\tilde{\lambda} = 1 \times 10^{-3}$, and $\tilde{T} = 1 \times 10^4$.}
\label{ext_damping}
\end{figure}

We plot the non-Poissonian part of the symmetrized-in-frequency current noise spectrum in units of the average current, i.e. $\Delta \bar{S}_I(\omega)/2e \langle I \rangle$, versus dimensionless frequency $\tilde{\omega}$ for different values of the dimensionless parameters.  First, we explore the regime of high bias voltage and high external temperature.  Each of Figs.~\ref{tunneling} - \ref{ext_damping} shows plots of the non-Poissonian current noise for different values of a certain dimensionless parameter, the other parameters being held fixed. The parameters being varied are the bare tunneling amplitude $t_0$ (Fig.~\ref{tunneling}), the bias voltage $\tilde{V}$ (Fig.~\ref{voltage}), the coupling $\tilde{\lambda}$ (Fig.~\ref{coupling}), the external temperature $\tilde{T}$ (Fig.~\ref{temp}), and the external oscillator damping $\tilde{\Gamma}_0$ (Fig.~\ref{ext_damping}).

In all the plots, one observes three peaks in the noise spectrum at $\tilde{\omega} = 0$, $\pm 2$, as well as two resonance-antiresonance features at $\tilde{\omega} = \pm 1$ (the noise is a symmetric function of $\tilde{\omega}$ and is not plotted for $\tilde{\omega} < 0$).  The peaks, especially the ones at $\tilde{\omega} = 0$, $\pm 2$, are very sharp and can be as high as $10^6$ for some parameter values, so it was necessary to truncate them in order to resolve the off-peak behavior.  To give a general sense of the peak magnitudes and their variation, the $\tilde{\omega} = 0$ peak ranges from $6 \times 10^2$ to $8 \times 10^5$, the $\tilde{\omega} = \pm 2$ peaks are usually half as high and range from $3 \times 10^2$ to $4 \times 10^5$, the $\tilde{\omega} = \pm 1$ resonance peaks range from $2 \times 10^{-3}$ to $1 \times 10^2$, and the antiresonance features range from $-1 \times 10^{-2}$ to $-2 \times 10^{-6}$.  The peaks and resonance-antiresonance features tend to broaden and become higher/deeper as $t_0$, $\tilde{V}$ and $\tilde{\lambda}$ increase (Figs.~\ref{tunneling} - \ref{coupling}).  The peaks at $\tilde{\omega} = 0$, $\pm 2$ broaden and become higher as $\tilde{T}$ increases, but broaden and become lower as $\tilde{\Gamma}_0$ increases.  At $\tilde{\omega} = \pm 1$, the resonance peak gets higher as $\tilde{T}$ increases and lower as $\tilde{\Gamma}_0$ increases, whereas the antiresonance gets shallower in both cases, and there is no noticeable change in the width of either feature (Figs.~\ref{temp} and~\ref{ext_damping}). 

Our noise spectrum is similar to that derived in Ref.~[\onlinecite{clerk04}] for a position-coupled oscillator and QPC with $\eta = 0$, except that we see a resonance-antiresonance feature instead of a positive peak at $\tilde{\omega} = \pm 1$.  In the $\eta = 0$ case, the noise spectrum near $\tilde{\omega} = \pm 1$ is proportional to the position spectrum of the oscillator.~\cite{clerk04}  It can be expressed as a leading term, corresponding to a classically fluctuating junction conductance, minus a quantum correction term, which arises from the correlations between the intrinsic shot noise of the detector and the back-action force on the oscillator.  This quantum correction is always smaller than the leading term when $\eta = 0$, resulting in a positive Lorentzian peak at $\tilde{\omega} = \pm 1$.  In Ref.~[\onlinecite{bennett08}], the non-Gaussian correlations between the junction current and back-action force are derived using a simple model, in which tunneling electrons impart random momentum kicks to the oscillator at the exact moment of tunneling, the typical size of the kicks being set by the Heisenberg uncertainty principle.  In the $\eta = 0$ case, the most notable effects of these correlations are an enhancement of the $\tilde{\omega} = 0$ peak and a suppression of the $\tilde{\omega} = \pm 2$ peaks relative to the classical picture.  However, in the $\eta = -\pi/2$ case they have a much more profound effect on the noise spectrum.  Ref.~[\onlinecite{doiron08}] shows that in this case the current noise near $\tilde{\omega} = \pm 1$ is proportional to the momentum spectrum of the oscillator (for our momentum-coupled system, this would be the position spectrum under the canonical transformation), and again there is a leading classical term and a quantum correction, but now the quantum correction can be larger than the classical term for a cold enough environment ($k_B T \ll eV$), producing a negative peak at $\tilde{\omega} = \pm 1$.  Even more interestingly, there is one more term in the noise at $\eta = -\pi/2$, namely the last term in Eq.~(8) of Ref.~[\onlinecite{doiron08}], which is non-negligible when the total oscillator damping due to the environment and the detector is very small, as is the case in our plots.  This is exactly the term leading to the Fano-like resonance-antiresonance features in our spectra.  Thus, due to the non-zero tunneling phase, our spectra show clear signatures of the correlations between the junction current and back-action force on the oscillator. 
 
It is also interesting to note that our spectra are similar to those obtained classically by Armour for the full noise spectrum of a SET whose capacitance depends linearly on the position of a nearby nanomechanical oscillator.~\cite{armour04}  In both cases, the same peaks and resonance-antiresonance features emerge, and the dependence of the peak heights and widths on the system parameters is very similar.  It is possible that the quantum effects in our system somehow mimic the classical effects in the SET-oscillator system.  The resonance-antiresonance features at $\tilde{\omega} = \pm 1$ are also predicted to appear in the back-action force spectrum for an oscillator coupled linearly via its position to a generic detector.~\cite{rodrigues09}  In our case, due to $\eta = -\pi/2$, the detector current is exactly correlated with the back-action force, so it is not surprising that we see the same features in the current noise spectrum as well. 

\begin{figure}[htbp]
\begin{center}
\includegraphics[width=0.75\textwidth]{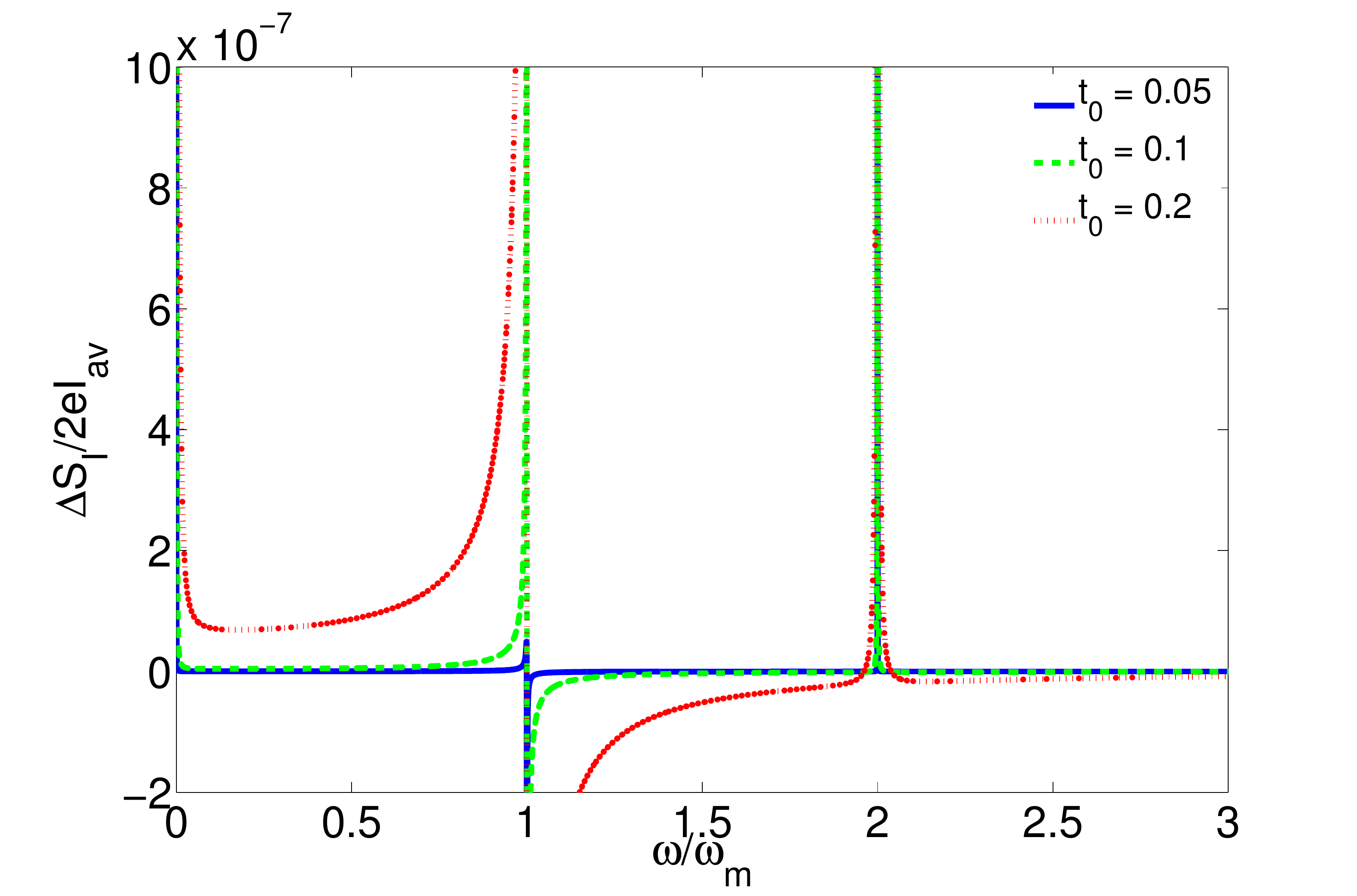}
\end{center}
\caption{Dimensionless non-Poissonian current noise spectrum for $t_0 = 0.05$, $0.1$ and $0.2$.  The remaining parameters are $\tilde{V} = 100$, $\tilde{\lambda} = 0.01$, $\tilde{T} = 0.01$ and $\tilde{\Gamma}_0 = 10^{-6}$.}
\label{tunneling_2}
\end{figure}

\begin{figure}[htbp]
\begin{center}
\includegraphics[width=0.75\textwidth]{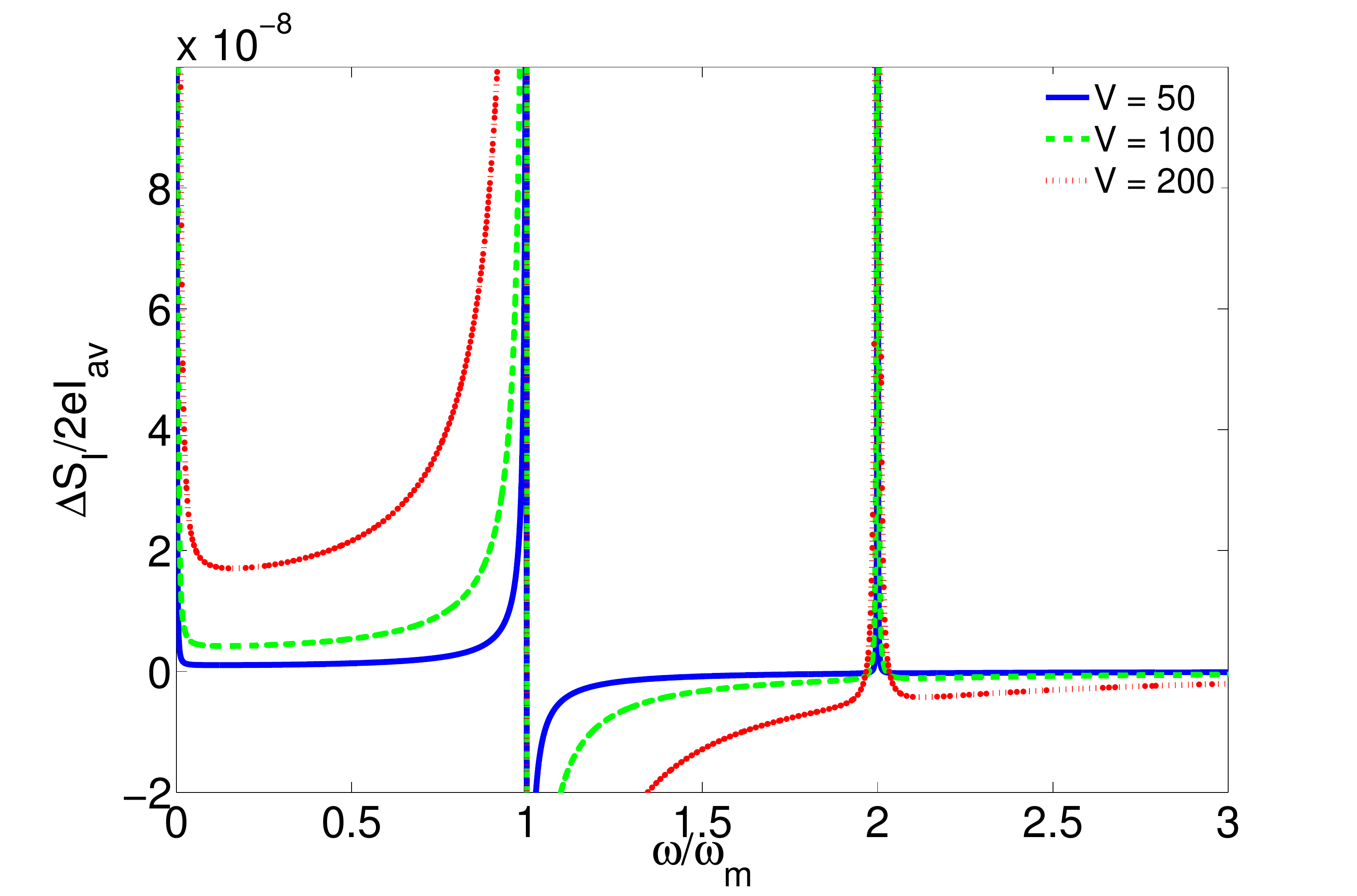}
\end{center}
\caption{Dimensionless non-Poissonian current noise spectrum for $\tilde{V} = 50$, $100$ and $200$.  The remaining parameters are $t_0 = 0.1$, $\tilde{\lambda} = 0.01$, $\tilde{T} = 0.01$ and $\tilde{\Gamma}_0 = 10^{-6}$.}
\label{voltage_2}
\end{figure}

\begin{figure}[htbp]
\begin{center}
\includegraphics[width=0.75\textwidth]{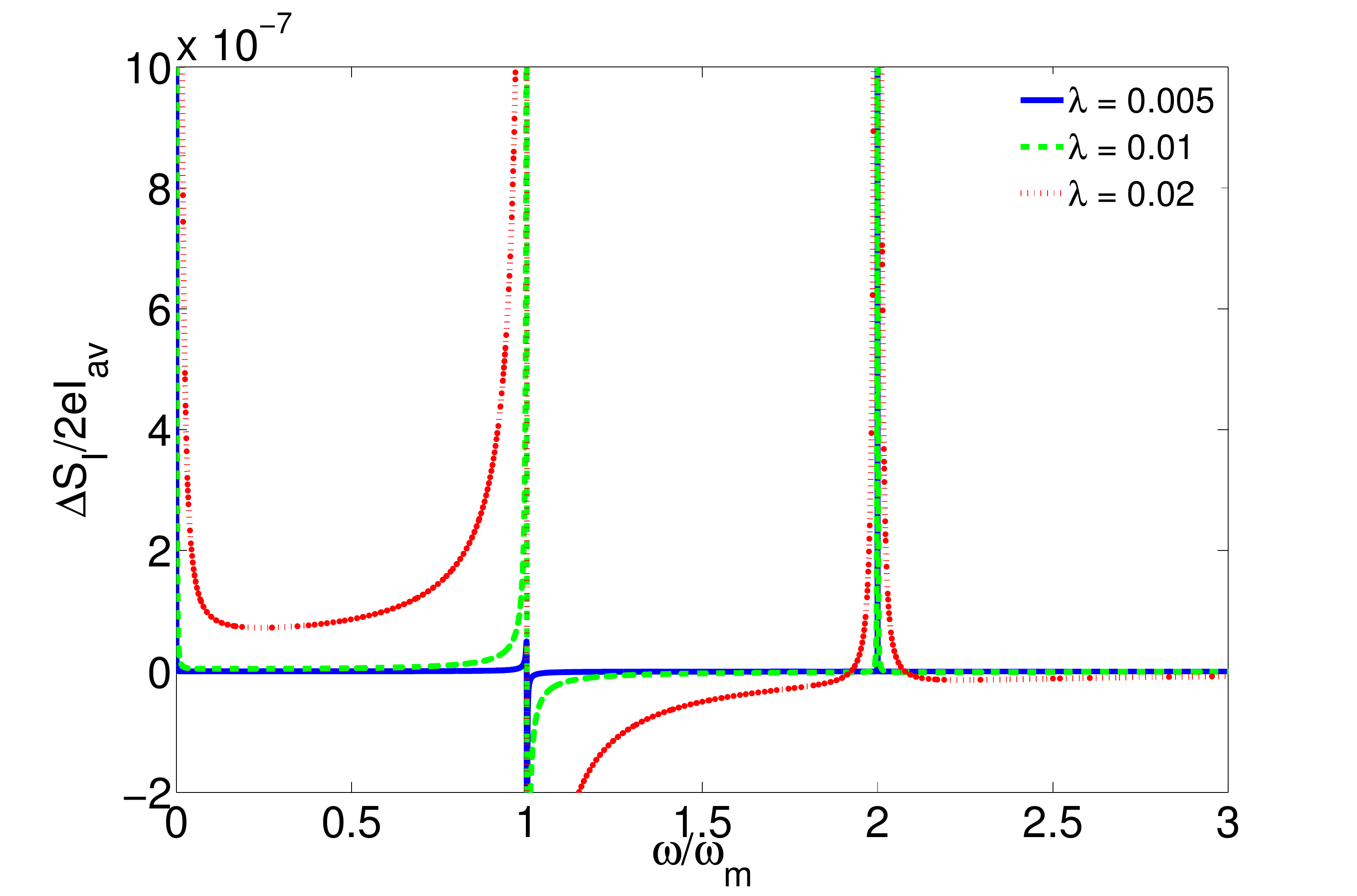}
\end{center}
\caption{Dimensionless non-Poissonian current noise spectrum for $\tilde{\lambda} = 0.005$, $0.01$ and $0.02$.  The remaining parameters are $t_0 = 0.1$, $\tilde{V} = 100$, $\tilde{T} = 0.01$ and $\tilde{\Gamma}_0 = 10^{-6}$.}
\label{coupling_2}
\end{figure}

\begin{figure}[htbp]
\begin{center}
\includegraphics[width=0.75\textwidth]{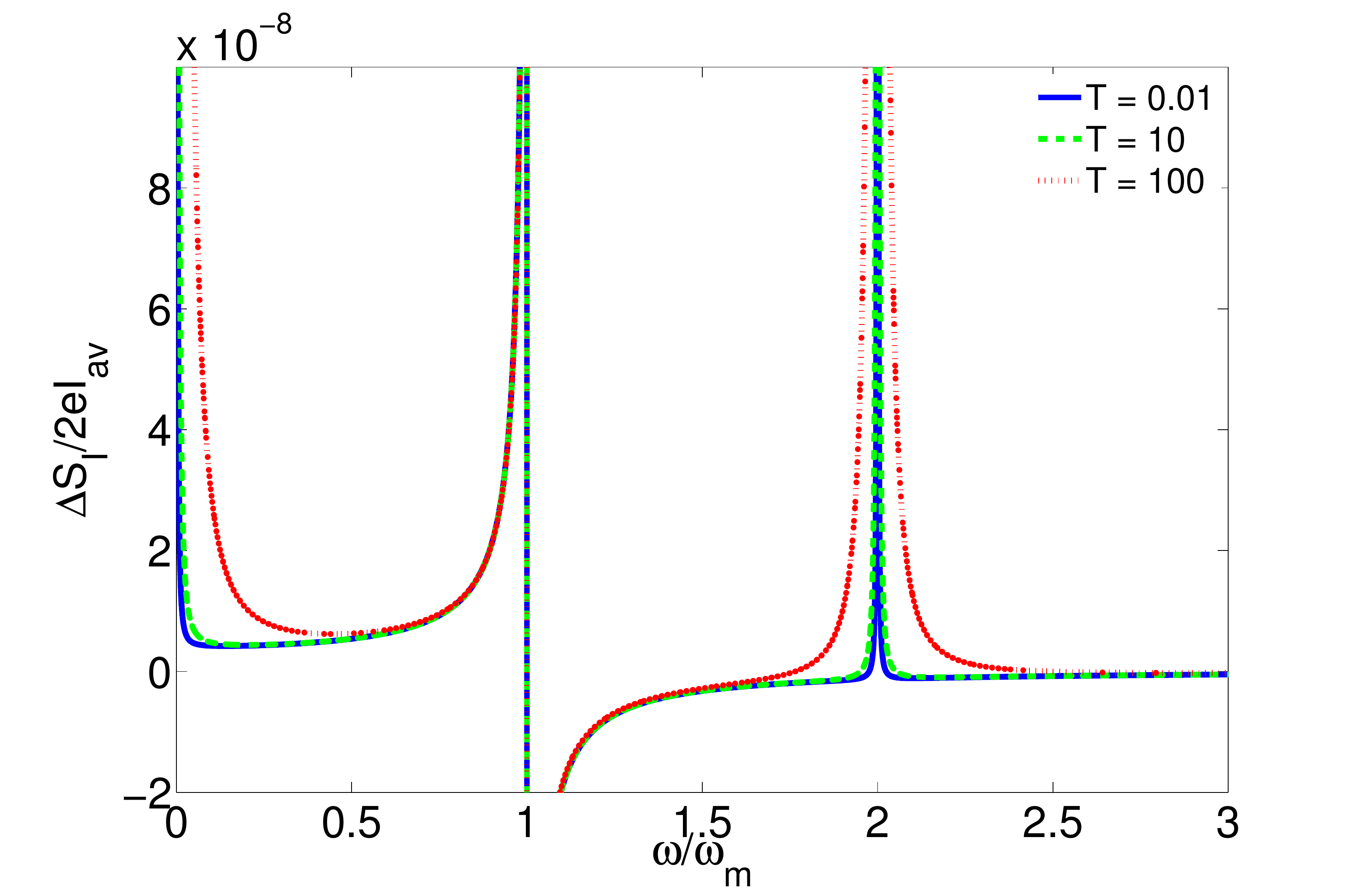}
\end{center}
\caption{Dimensionless non-Poissonian current noise spectrum for $\tilde{T} = 0.01$, $10$ and $100$.  The remaining parameters are $t_0 = 0.1$, $\tilde{V} = 100$, $\tilde{\lambda} = 0.01$, and $\tilde{\Gamma}_0 = 10^{-6}$.}
\label{temp_2}
\end{figure}

\begin{figure}[htbp]
\begin{center}
\includegraphics[width=0.75\textwidth]{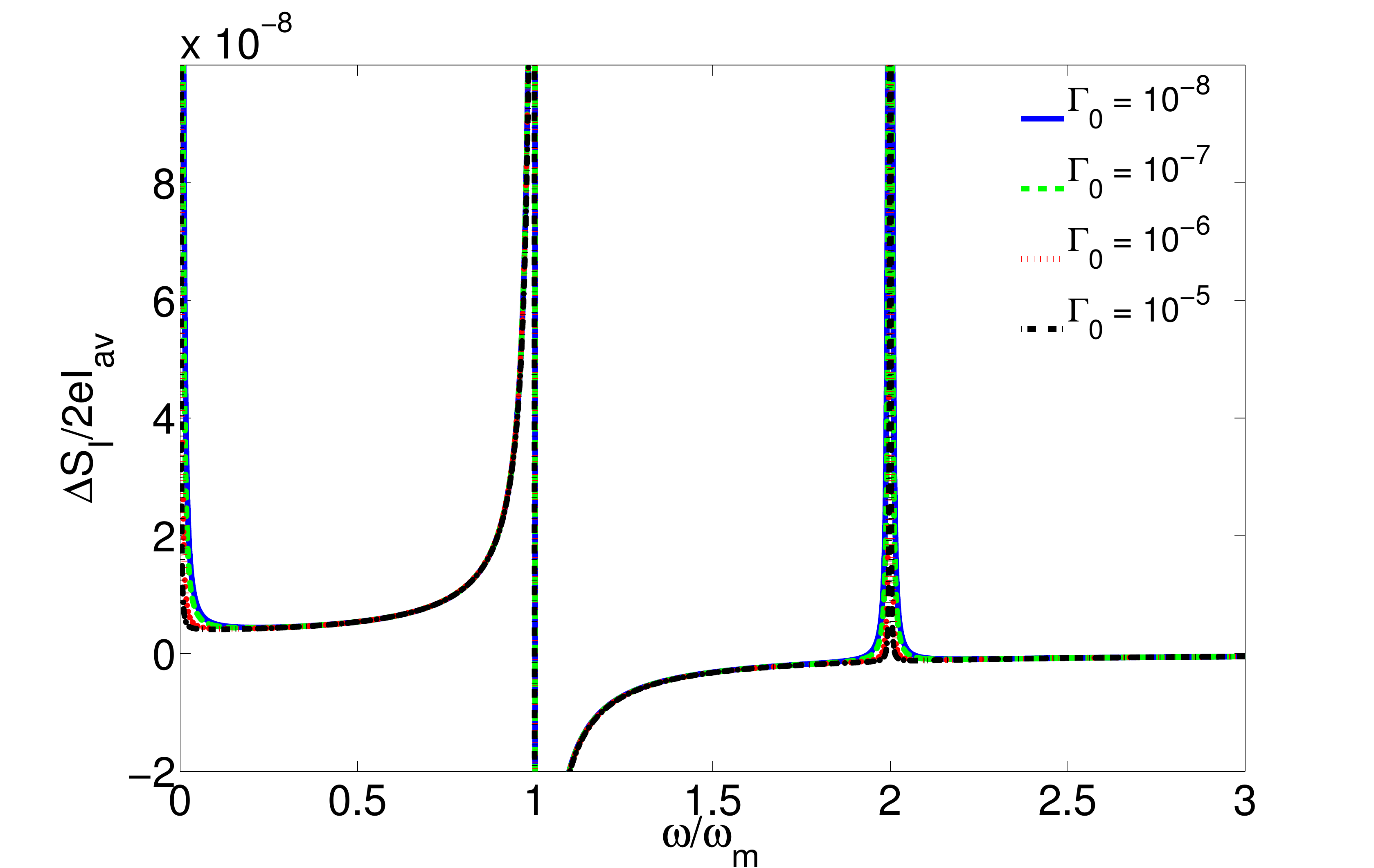}
\end{center}
\caption{Dimensionless non-Poissonian current noise spectrum for $\tilde{\Gamma}_0 = 10^{-8}$, $10^{-7}$, $10^{-6}$ and $10^{-5}$.  The remaining parameters  are $t_0 = 0.1$, $\tilde{V} = 100$, $\tilde{\lambda} = 0.01$, and $\tilde{T} = 0.01$.}
\label{ext_damping_2}
\end{figure}

Next, we focus on the regime of low external temperature, relatively low bias voltage (but still $\tilde{V} \gg 1$, as required by the Born-Markov approximation), and comparable external and internal damping, i.e. $\tilde{\Gamma}_0 \approx t_0^2 \tilde{\lambda}^2 / 2 \pi$.  We start with $t_0 = 0.1$, $\tilde{V} = 100$, $\tilde{\lambda} = 0.01$, $\tilde{T} = 0.01$ and $\tilde{\Gamma}_0 = 10^{-6}$ as the central point in our parameter space, and vary each parameter around its central value, keeping the other parameters fixed (Figs.~\ref{tunneling_2} - \ref{ext_damping_2}).  In Fig.~\ref{temp_2}, the spectrum does not change appreciably as a function of external temperature for $\tilde{T} \leq 1$, hence it is only plotted for $\tilde{T} = 0.01$ and $\tilde{T} \geq 10$.  In this regime, the magnitude of the $\tilde{\omega} = 0$ peak ranges from $9 \times 10^{-4}$ to $80$, the $\tilde{\omega} = \pm 2$ peaks range from $6 \times 10^{-4}$ to $40$, the $\tilde{\omega} = \pm 1$ resonances vary between $5 \times 10^{-8}$ and $2 \times 10^{-4}$, and the antiresonance features vary from $-1$ to $-2 \times 10^{-2}$.  The dependence of the peak heights and widths on the parameters is much the same as in the high-voltage, high-temperature regime.  The main difference between the two regimes is that the $\tilde{\omega} = \pm 1$ antiresonance features are relatively more prominent in the low-temperature regime.  In fact, the set of parameter values $t_0 = 0.1$, $\tilde{V} = 100$, $\tilde{\lambda} = 0.01$, $\tilde{T} = 0.01$ and $\tilde{\Gamma}_0 = 10^{-6}$ appears to be a crossover point in parameter space, where all the peaks are similar in magnitude (e.g.~the $\tilde{\omega} = 0$ peak magnitude is $0.3$, the $\tilde{\omega} = \pm 2$ peaks are $0.15$, and the $\tilde{\omega} = \pm 1$ antiresonances are $-0.25$; the resonances are negligible throughout this regime).  For higher values of $t_0$, $\tilde{V}$, $\tilde{\lambda}$ and $\tilde{T}$, and lower values of $\tilde{\Gamma}_0$, the $\tilde{\omega} = 0$, $\pm 2$ peaks dominate, and for lower values of $t_0$, $\tilde{V}$, $\tilde{\lambda}$ and $\tilde{T}$, and higher values of $\tilde{\Gamma}_0$, the $\tilde{\omega} = \pm 1$ antiresonances dominate.  Besides, unlike in the high-temperature, high-voltage regime, here the current noise spectrum is sub-Poissonian ($\Delta \bar{S}_I < 0$) for $|\tilde{\omega}| > 1$, except near the $\tilde{\omega} = \pm 2$ peaks, especially for large $t_0$, $\tilde{V}$ or $\tilde{\lambda}$, and approaches zero from below as $\tilde{\omega} \rightarrow \pm \infty$.  This is consistent with Ref.~[\onlinecite{doiron08}], where the last term in Eq.~(8), which is responsible for the Fano-like feature at $\tilde{\omega} = \pm 1$, grows with $t_0$, $\tilde{V}$ and $\tilde{\lambda}$.  Also, the first (Lorentzian) term is expected to be negative when $k_B T \ll eV$, explaining the suppression of the resonance peak in this regime. 

\begin{figure}[htbp]
\begin{center}
\includegraphics[width=0.75\textwidth]{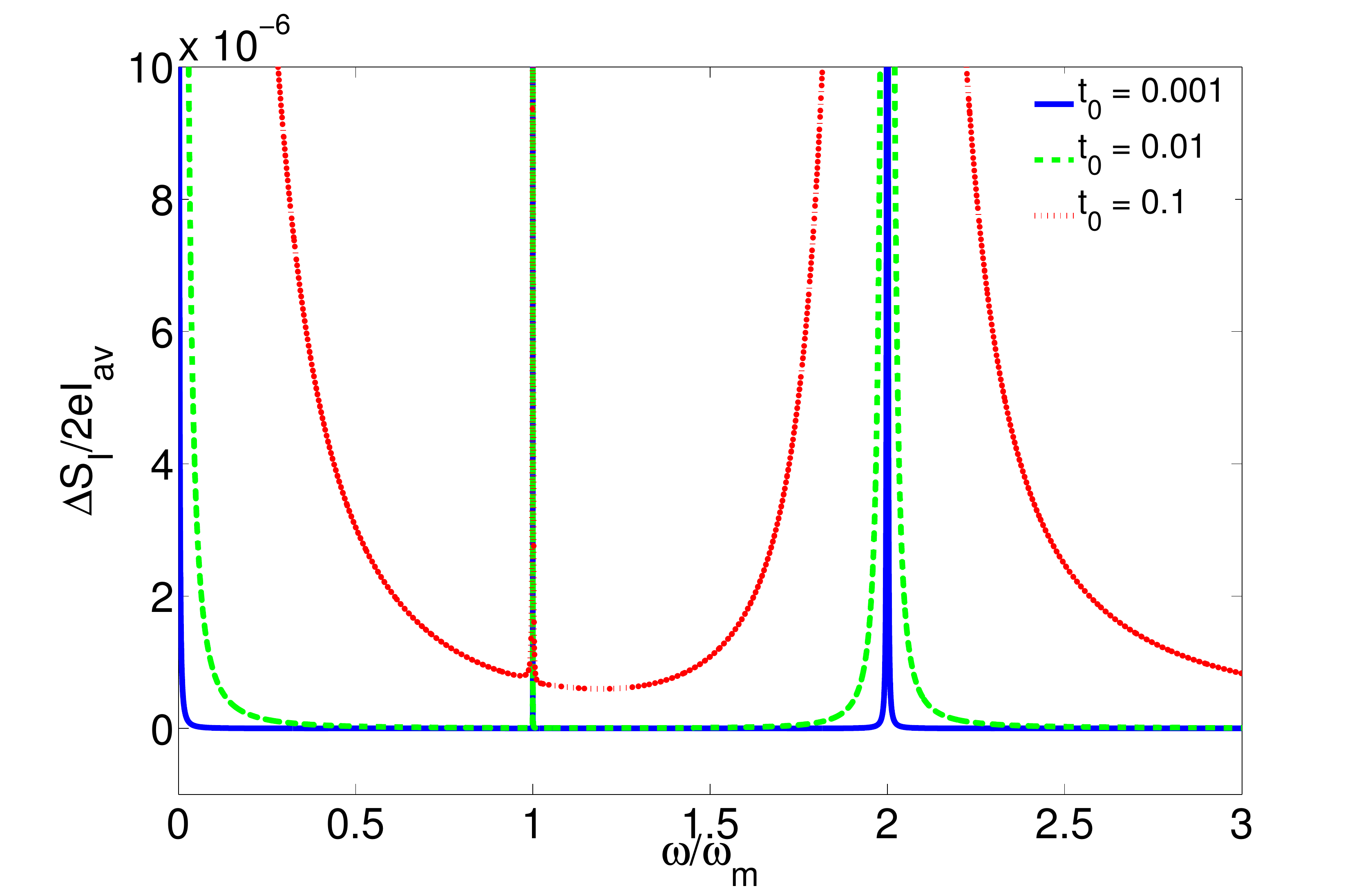}
\end{center}
\caption{Dimensionless non-Poissonian current noise spectrum for $t_0 = 0.001$, $0.01$ and $0.1$.  The remaining parameters are $\tilde{V} = 100$, $\tilde{\lambda} = 0.01$, $\tilde{T} = 10^4$ and $\tilde{\Gamma}_0 = 10^{-6}$.}
\label{tunneling_high_T}
\end{figure}

\begin{figure}[htbp]
\begin{center}
\includegraphics[width=0.75\textwidth]{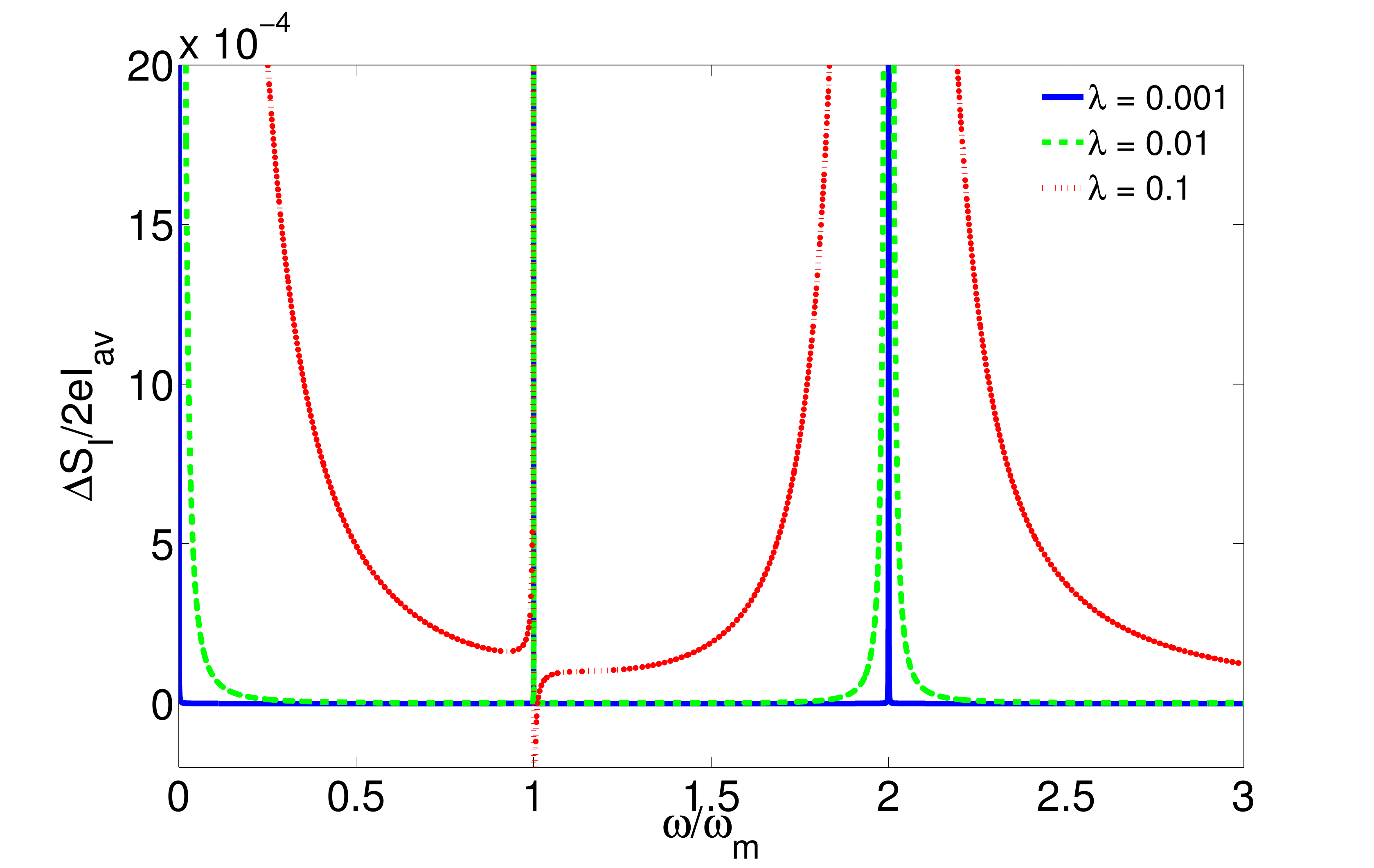}
\end{center}
\caption{Dimensionless non-Poissonian current noise spectrum for $\tilde{\lambda} = 0.001$, $0.01$ and $0.1$.  The remaining parameters are $t_0 = 0.1$, $\tilde{V} = 100$, $\tilde{T} = 10^4$ and $\tilde{\Gamma}_0 = 10^{-6}$.}
\label{coupling_high_T}
\end{figure}

\begin{figure}[htbp]
\begin{center}
\includegraphics[width=0.75\textwidth]{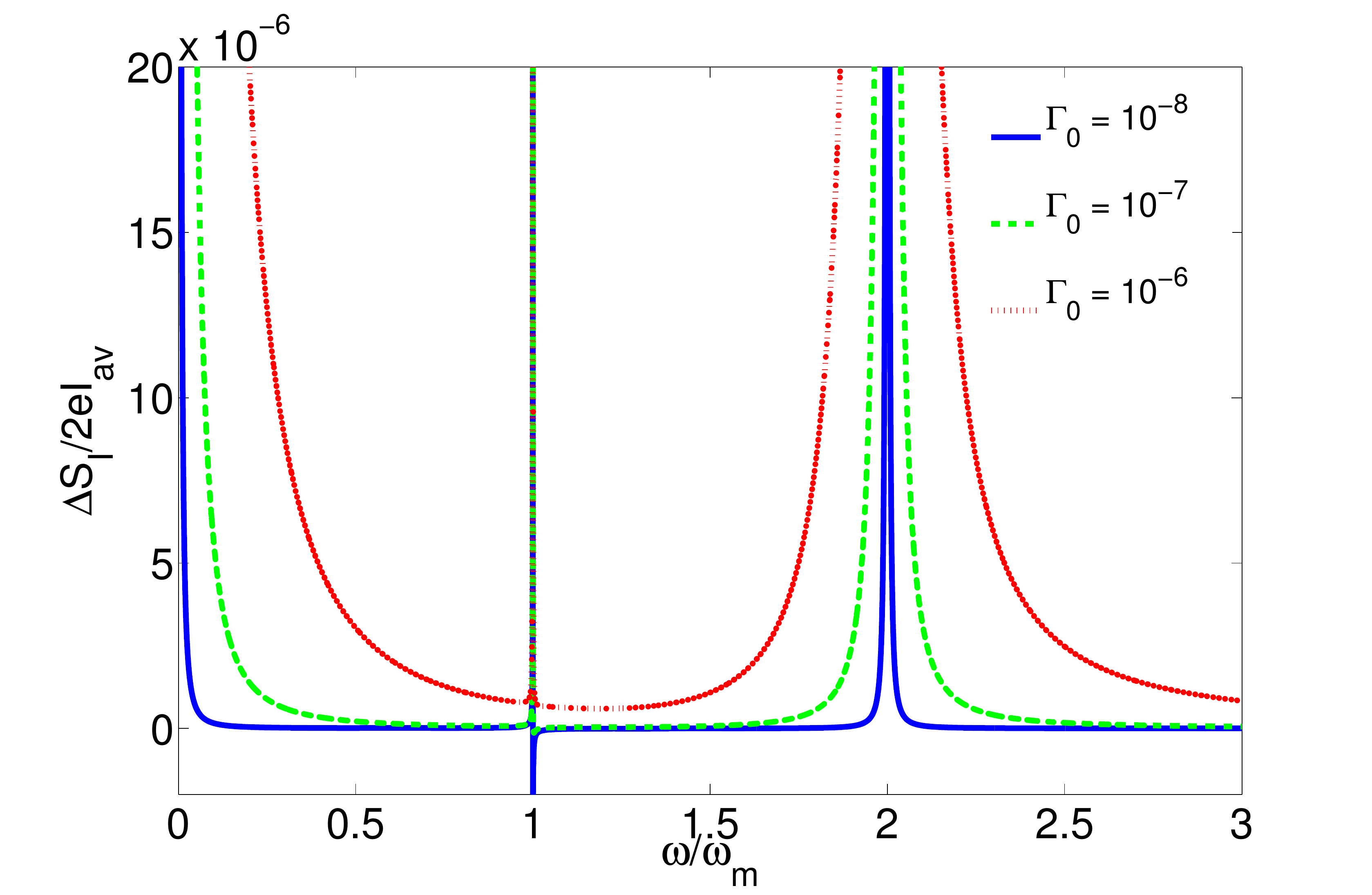}
\end{center}
\caption{Dimensionless non-Poissonian current noise spectrum for $\tilde{\Gamma}_0 = 10^{-8}$, $10^{-7}$ and $10^{-6}$.  The remaining parameters are $t_0 = 0.1$, $\tilde{V} = 100$, $\tilde{\lambda} = 0.01$ and $\tilde{T} = 10^4$.}
\label{ext_damping_high_T}
\end{figure}

Finally, we investigate the regime of high external temperature ($\tilde{T} = 10^4$) and relatively low bias voltage ($\tilde{V} = 10^2$), and vary each of the remaining three parameters around the parameter-space point with coordinates $t_0 = 0.1$, $\tilde{\lambda} = 0.01$ and $\tilde{\Gamma}_0 = 10^{-6}$, keeping the other two parameters fixed (Figs.~\ref{tunneling_high_T} - \ref{ext_damping_high_T}).  We find good qualitative agreement between this regime and the high-voltage, high-temperature one presented in Figs.~\ref{tunneling} - \ref{ext_damping}.  The $\tilde{\omega} = 0$, $\pm 2$ peaks are again much higher and broader than the $\tilde{\omega} = \pm 1$ resonance-antiresonance features, and the magnitudes of the peaks and their dependence on the parameters are very similar in both regimes.  This agreement is not surprising given that the two regimes have very similar values for all parameters except the voltage.  Apparently, simply lowering the bias voltage, while keeping it large compared to the oscillator frequency, does not introduce us to a fundamentally new regime.  Nevertheless, there are some subtle differences.  First, in the low-voltage regime the antiresonance features are even less pronounced and disappear completely for most parameter values, as seen in the figures.  This is expected, as a high external temperature can wash out the noise suppression due to a weak (low bias voltage) detector, especially at weak coupling, weak tunneling and/or high external damping.  Secondly, there are slight departures from monotonicity in the dependence of the peak heights on the coupling and external damping parameters.  The peak heights actually $decrease$ as $\tilde{\lambda}$ increases from $0.01$ to $0.1$, and $increase$ as $\tilde{\Gamma}_0$ increases up to $10^{-7}$ before eventually decreasing.  At present, it is not quite clear what causes these slight aberrations.

In the experiment of Ref.~[\onlinecite{stettenheim10}], the parameter values (e.g. for sample A) are $\tilde{V} \approx 10^4$, $\tilde{\lambda} = 7.2 \times 10^{-6}$, $\tilde{\Gamma}_0 = 3.3 \times 10^{-2}$, and $\tilde{T} = 1.6 \times 10^3$,~\cite{stettenheim10} corresponding to our high-voltage, high-temperature regime, with the external damping much larger than the detector damping ($\tilde{\Gamma}_0 \gg t_0^2 \tilde{\lambda}^2 / 2 \pi$).  It is important to note, however, that the tunneling $t_0$ is much larger in the experiment ($t_0^2 = 0.5$).  Comparing the experimental and theoretical results (e.g. Fig. 4d in Ref.~[\onlinecite{stettenheim10}] and Figs.~\ref{tunneling} - \ref{ext_damping} in this article), we find the same peaks at $\tilde{\omega} = 0$, $\pm 1$ and $\pm 2$.  Both the experimental Fano factor and the theoretical current noise vary over many orders of magnitude, indicating strong electron-electron correlations due to interaction with the oscillator.  The $\tilde{\omega} = \pm 2$ peaks are somewhat less pronounced in the experimental results.  The small antiresonances predicted theoretically in this regime are not resolved in the experiment, possibly due to the background noise.  The sub-Poissonian noise observed at higher frequencies in the experiment is absent from Figs.~\ref{tunneling} - \ref{ext_damping}, but surprisingly does appear in the low-temperature, low-voltage regime in Figs.~\ref{tunneling_2} - \ref{ext_damping_2}.

\begin{figure}[htbp]
\begin{center}
\includegraphics[width=0.75\textwidth]{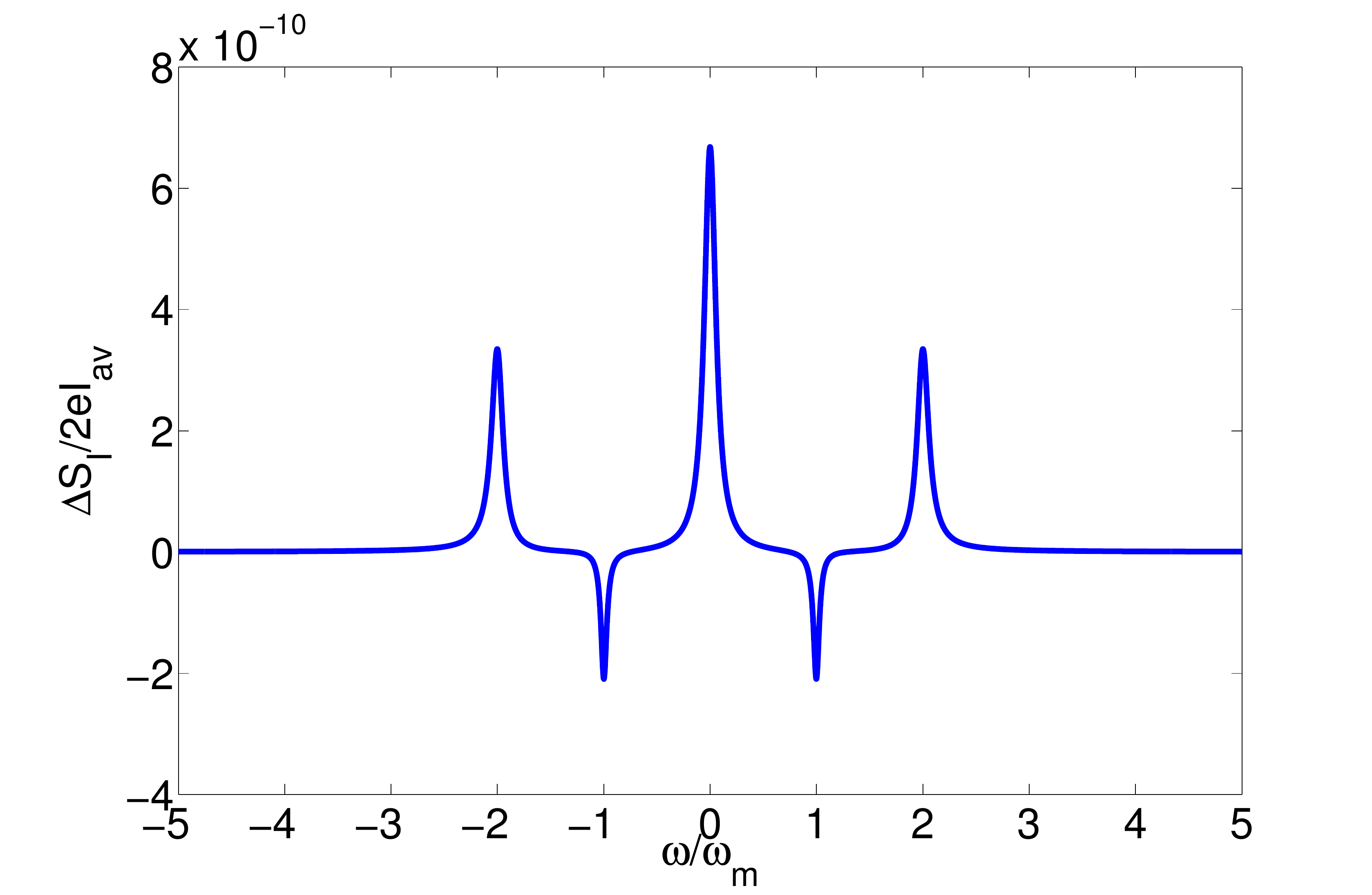}
\end{center}
\caption{Dimensionless non-Poissonian current noise spectrum for the parameter values used for sample A in the experiment of Ref.~[\onlinecite{stettenheim10}]: $t_0 = 0.71$, $\tilde{V} = 1 \times 10^4$, $\tilde{\lambda} = 7.2 \times 10^{-6}$, $\tilde{T} = 1.6 \times 10^3$, and $\tilde{\Gamma}_0 = 3.3 \times 10^{-2}$.}
\label{expcomp}
\end{figure}

For comparison purposes, in Fig.~\ref{expcomp} we have also plotted the theoretical current noise spectrum for the exact set of parameter values used in the experiment.  The theory fails to predict the high super-Poissonian peaks, the resonances at $\tilde{\omega} = \pm 1$ or the sub-Poissonian noise seen in the experiment.  It also overestimates the relative magnitude of the $\tilde{\omega} = \pm 2$ peaks.  These discrepancies are probably due to the breakdown of the theoretical method at strong tunneling, and suggest that comparisons between theory and experiment should be made with caution until the strong tunneling regime of our system has been investigated theoretically.  As expected, our theory fails to predict the sharp increase in peak heights as the system enters the strong tunneling regime.  In fact, making the coupling larger and the external damping smaller than in the experiment, as in Figs.~\ref{tunneling} - \ref{ext_damping}, partially \textquotedblleft compensates\textquotedblright for this failure, producing higher peaks more similar to those seen in the experiment. 

\section{\label{sec:steadystate}Wigner Function Representation and Steady-State Oscillator Dynamics}
In order to investigate the steady-state behavior of the oscillator, it is useful to translate the master equation into the Wigner-Weyl formalism.~\cite{case08}  Using the method outlined in Bennett,~\cite{bennett08} one obtains:
\begin{eqnarray}
\partial_t W(x, p; \chi; t) & = & \bigg\{ -\frac{p}{m}\partial_x + m \omega_m^2 x \partial_p + \bar{F}_0\bigg(-\frac{\pi}{2} \bigg) \partial_x \nonumber \\
& + & D_+ \partial_x^2 + 2 m^2 \omega_m^2 \tilde{\gamma}_+ \partial_x  x + D_0 \partial_p^2 + 2 \tilde{\gamma}_0 \partial_p p \nonumber \\
& + & \big( e^{i \chi} - 1 \big) \bigg[\Gamma_+(0) - \frac{\hbar}{2} \Gamma_+(0) \frac{t_1}{t_0} \partial_x - \frac{D_+}{\hbar} \frac{t_0}{t_1} \partial_x + \frac{2 D_+}{\hbar^2} \bigg(p^2 + \frac{\hbar^2}{4} \partial_x^2 \bigg) \nonumber \\
& - & \frac{2 m^2 \omega_m^2 \tilde{\gamma}_+}{\hbar} \frac{t_0}{t_1} x + m^2 \omega_m^2 \tilde{\gamma}_+ \big(p \partial_p + \partial_x x \big) \bigg] \bigg\} W(x, p; \chi; t).
\label{Wigner}
\end{eqnarray}
Here we have assumed large bias voltage (the $\sigma = -1$ terms are set to zero).  In what follows, we also assume zero temperature in the leads, just as in the previous sections.  Just as in the case of the position-coupled oscillator and QPC studied in Ref.~[\onlinecite{bennett08}], a careful inspection of the above equation shows that exactly half of the back-action damping and diffusion (terms involving $\tilde{\gamma}_+$ and  $D_+$ on the second line of Eq.~(\ref{Wigner}) as well as the very last terms on the third and fourth lines) is correlated with tunneling (multiplied by $e^{i \chi}$), whereas the other half is independent of tunneling events.  The precise correlation between the electron tunneling events and the momentum kicks imparted to the oscillator suggests that there is a departure from the simple model of the detector as a thermal bath.  On the other hand, the uncorrelated half of the back-action means that one gains information about the oscillator even when no electrons are tunneling. 

The first two lines of Eq.~(\ref{Wigner}) represent a classical Fokker-Planck equation for an oscillator coupled to two equilibrium baths, and agree well with Eq.~(6a) in Ref.~[\onlinecite{bennett08}], except that in our case the external bath is position-coupled whereas the detector bath is momentum-coupled.  Besides, the average back-action force on the oscillator, $\bar{F}_0(-\pi/2)$, is non-zero in our case.  The first four terms on the third line of Eq.~(\ref{Wigner}) combine to give an oscillator-dependent tunneling rate through the QPC, analogous to the classically fluctuating tunneling rate represented by the last term of Eq.~(6a) in Ref.~[\onlinecite{bennett08}].  However, there are subtle differences - instead of a quadratic dependence of the classical tunneling rate on the oscillator position, in our case there is a quadratic dependence on the oscillator momentum, as well as a linear term proportional to $\partial_x W$.  Finally, the first two terms on the fourth line of Eq.~(\ref{Wigner}) correspond to the last two terms in Eq.~(6b) of Ref.~[\onlinecite{bennett08}], which represent quantum corrections to the average tunneling rate and arise from the difference between tunneling processes involving absorption or emission of a phonon.  There are again some differences - in the position-coupled system, these terms involve $\partial_x W$ and $x \partial_x W$, whereas in our case they are proportional to $x W$ and $p \partial_p W$.

To study the steady-state dynamics of the oscillator, we need to trace over $N$, the number of electrons that have tunneled through the detector, which is equivalent to setting $\chi = 0$ in the above equation (cf. Eq.~(\ref{NchiFT})).  The resulting simpler equation can be integrated by parts to yield coupled equations for the oscillator moments, e.g. up to second order:
\begin{eqnarray}
\frac{d}{dt} \langle x \rangle & = & \frac{\hbar t_0 t_1 eV}{h} -2 m^2 \omega_m^2 \tilde{\gamma}_+ \langle x \rangle + \frac{1}{m} \langle p \rangle = 0 \nonumber \\
\frac{d}{dt} \langle p \rangle & = & -m \omega_m^2 \langle x \rangle - 2 \tilde{\gamma}_0 \langle p \rangle = 0 \nonumber \\
\frac{d}{dt} \langle x^2 \rangle & = & \frac{\hbar^2 t_1^2 eV}{h} + \frac{2 \hbar t_0 t_1 eV}{h} \langle x \rangle - 4 m^2 \omega_m^2 \tilde{\gamma}_+ \langle x^2 \rangle + \frac{2}{m} \langle xp \rangle = 0 \nonumber \\  
\frac{d}{dt} \langle xp \rangle & = & \frac{\hbar t_0 t_1 eV}{h} \langle p \rangle - m \omega_m^2 \langle x^2 \rangle -2 \big(m^2 \omega_m^2 \tilde{\gamma}_+ + \tilde{\gamma}_0 \big) \langle xp \rangle + \frac{1}{m} \langle p^2 \rangle = 0 \nonumber \\
\frac{d}{dt} \langle p^2 \rangle & = & 2 D_0 -2 m \omega_m^2 \langle xp \rangle - 4 \tilde{\gamma}_0 \langle p^2 \rangle = 0
\end{eqnarray}
As expected, these equations are the same as Eqs.~(\ref{firstordermoments}) and~(\ref{secondordermoments}) in Appendix~\ref{sec:appendixc}.  In the above equations, $\langle xp \rangle$ represents the symmetrized moment $\langle xp + px \rangle/2$, which is why the imaginary terms from Eqs.~(\ref{firstordermoments}) and~(\ref{secondordermoments}) are absent.  We will stick to this convention for the rest of this section.  It is interesting to consider the limiting cases in which either the detector or the environment decouples from the oscillator.  In the former case ($\tilde{\gamma}_+$, $V \rightarrow 0$), one obtains $\langle \langle xp \rangle \rangle = 0$, $\langle \langle x^2 \rangle \rangle = k_B T/m \omega_m^2$ and $\langle \langle p^2 \rangle \rangle = m k_B T$ (when $k_B T \gg \hbar \omega_m$), consistent with the equipartition theorem.  Note that we are working with the irreducible moments, i.e. the variance and covariance, hence the double brackets.  In the latter case ($\tilde{\gamma}_0$, $D_0 \rightarrow 0$), one also obtains equipartition results:  $\langle \langle xp \rangle \rangle = 0$, $\langle \langle x^2 \rangle \rangle = k_B T_{\mathrm{det}}/m \omega_m^2$ and $\langle \langle p^2 \rangle \rangle = m k_B T_{\mathrm{det}}$, where we have defined the detector temperature $T_{\mathrm{det}} = eV/2 k_B$.

\begin{figure}[htbp]
\begin{center}
\includegraphics[width=0.75\textwidth]{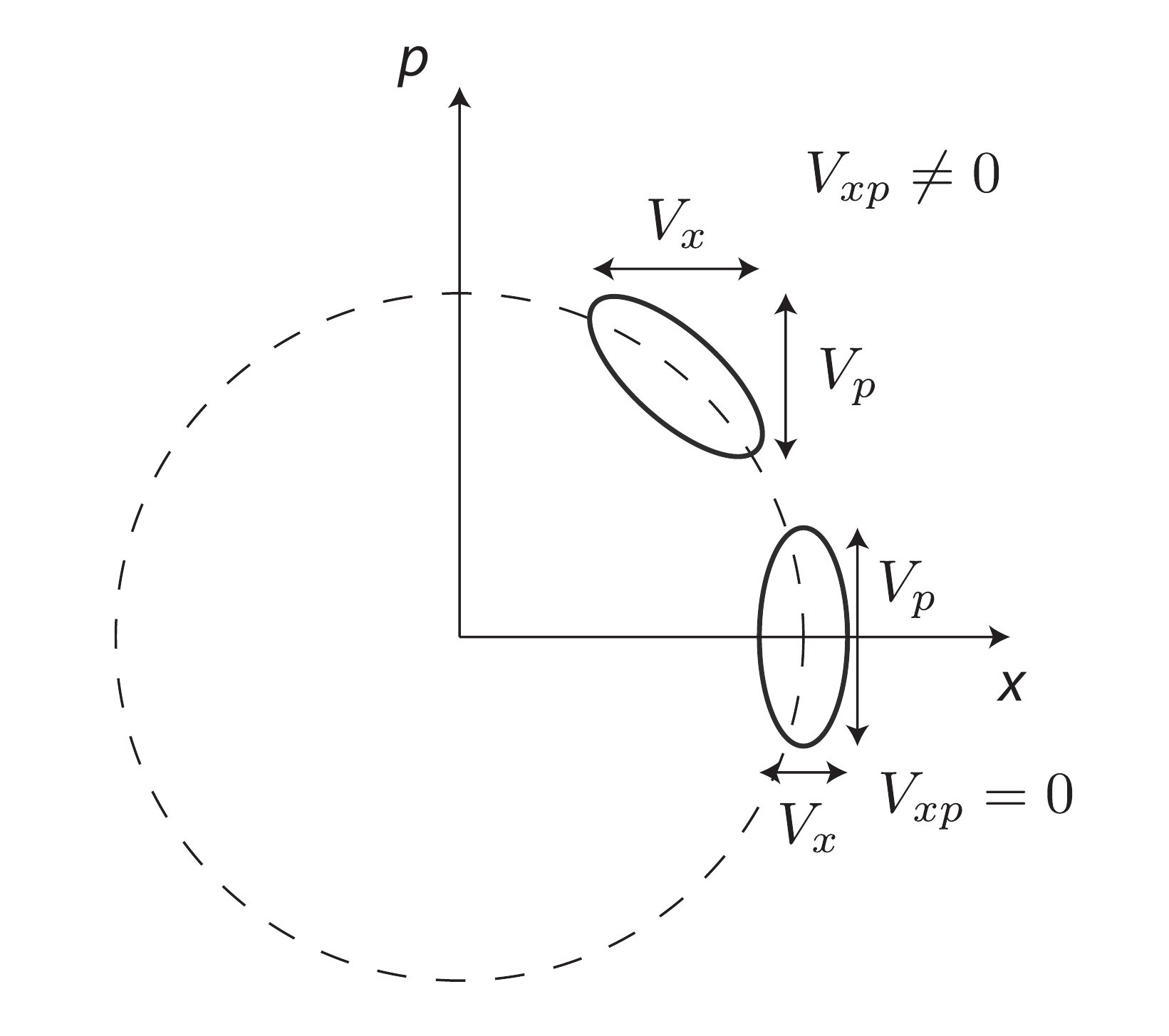}
\end{center}
\caption{Phase-space diagram of a squeezed state showing that one of the variances, $V_x$ or $V_p$, is minimized when the covariance $V_{xp}$ is zero.  The solid ellipses represent the Wigner function contours of a squeezed state in two different orientations relative to the axes.}
\label{squeezefig}
\end{figure}

In fact, one can solve the moment equations algebraically in the general case.  One obtains the following results:
\begin{eqnarray}
\langle x \rangle & = & \frac{2 \hbar t_0 t_1 eV \tilde{\gamma}_0}{h \omega_m^2 (4 m^2 \tilde{\gamma}_0 \tilde{\gamma}_+ + 1)}, \nonumber \\
\langle p \rangle & = & - \frac{m \hbar t_0 t_1 eV}{h (4m^2 \tilde{\gamma}_0 \tilde{\gamma}_+ + 1)}, \nonumber \\
\langle \langle x^2 \rangle \rangle & = & \frac{m^2 \hbar^2 t_1^2 eV (4 \tilde{\gamma}_0^2 + 4m^2 \omega_m^2 \tilde{\gamma}_0 \tilde{\gamma}_+ + \omega_m^2) + 2 h D_0}{4 m^2 \omega_m^2 h (4m^2 \tilde{\gamma}_0 \tilde{\gamma}_+ + 1)(\tilde{\gamma}_0 + m^2 \omega_m^2 \tilde{\gamma}_+)}, \nonumber \\
\langle \langle xp \rangle \rangle & = & - \frac{m(\tilde{\gamma}_0 \hbar^2 t_1^2 eV - 2 h \tilde{\gamma}_+ D_0)}{2 h (4 m^2 \tilde{\gamma}_0 \tilde{\gamma}_+ + 1)(\tilde{\gamma}_0 + m^2 \omega_m^2 \tilde{\gamma}_+)}, \nonumber \\
\langle \langle p^2 \rangle \rangle & = & \frac{2 h D_0 (4 m^2 \tilde{\gamma}_0 \tilde{\gamma}_+ + 4m^4 \omega_m^2 \tilde{\gamma}_+^2 + 1) + m^2 \omega_m^2 \hbar^2 t_1^2 eV}{4 h (4 m^2 \tilde{\gamma}_0 \tilde{\gamma}_+ + 1)(\tilde{\gamma}_0 + m^2 \omega_m^2 \tilde{\gamma}_+)}.
\label{moments}
\end{eqnarray}
The non-zero values of $\langle x \rangle$ and $\langle p \rangle$ imply that the oscillator is in a boosted frame as a result of our use of an effective model for the physical system of Ref.~[\onlinecite{stettenheim10}].  It is interesting to know if the simultaneous interaction of the oscillator with the position-coupled external bath and the momentum-coupled detector bath can put the oscillator into a so-called squeezed state - a clear signature of quantum behavior in the system.  If the initial state of the oscillator has a Gaussian Wigner function, the quadratic form of Eq.~(\ref{Wigner}) ensures that it remains Gaussian for all time.  Such a state will obey the position-momentum Heisenberg uncertainty relation, i.e. in dimensionless form
\begin{equation}
V_x V_p \ge 1,
\end{equation}
where $V_x = \langle \langle \tilde{x}^2 \rangle \rangle$ and $V_p = \langle \langle \tilde{p}^2 \rangle \rangle$ are the variances in dimensionless units.  For a squeezed state, we need either $V_x < 1$ or $V_p < 1$.~\cite{henry88}  The values of $V_x$ and $V_p$ depend on the orientation of the $x$ and $p$ axes in phase space.  Fig.~\ref{squeezefig} shows that $V_x$ or $V_p$ is minimized when the covariance $V_{xp} = \langle \langle \tilde{x} \tilde{p} + \tilde{p} \tilde{x} \rangle \rangle /2$ is zero.  Thus we need to find a new set of axes in phase space such that $V_{xp} = 0$, and therefore $V_x$ (or equivalently $V_p$) attains its minimum value.  Mathematically, this is analogous to diagonalizing the variance-covariance matrix of the Gaussian state.  If the smaller of the eigenvalues thus obtained is less than unity, we have a squeezed state.

Using this criterion, we don't find any evidence of quantum squeezing in our system in any of the parameter regimes discussed in the previous section.  The variances are larger than unity and approximately equal, and the covariance is much smaller than the variances, resulting in equal, larger-than-unity eigenvalues.  The absence of quantum squeezing is not surprising given the form of Eq.~(\ref{Wigner}).  As already discussed, when we set $\chi = 0$, only the first two lines of the equation remain, which represent a classical Fokker-Planck equation for an oscillator coupled to two independent reservoirs - the position-coupled environment and the momentum-coupled detector.  Thus an unconditional measurement of the steady state of the oscillator cannot be expected to yield any quantum results.  Only a measurement conditioned on a certain history of the current can show departures from the effective bath model due to the correlation between the tunneling and back-action already present in the master equation.~\cite{bennett08} 

A natural question to ask is whether the covariance can ever be non-negligible compared to the variances.  To answer this question fully, we look at the analytical expressions for the variances and covariance, i.e. the dimensionless versions of Eq.~(\ref{moments}):
\begin{eqnarray}
V_x & = & \frac{ \frac{1}{2 \pi} t_0^2 \tilde{\lambda}^2 \tilde{V} \big( 4 \tilde{\Gamma}_0^2 + \frac{2}{\pi} t_0^2 \tilde{\lambda}^2 \tilde{\Gamma}_0 + 1 \big) + \tilde{\Gamma}_0 \coth \big(\frac{1}{2 \tilde{T}} \big)}{\big(\frac{2}{\pi} t_0^2 \tilde{\lambda}^2 \tilde{\Gamma}_0 + 1 \big) \big(\tilde{\Gamma}_0 + \frac{1}{2 \pi} t_0^2 \tilde{\lambda}^2 \big)}, \nonumber \\
V_p & = & \frac{\tilde{\Gamma}_0 \coth \big( \frac{1}{2\tilde{T}} \big) \big(\frac{2}{\pi} t_0^2 \tilde{\lambda}^2 \tilde{\Gamma}_0 + \frac{1}{\pi^2} t_0^4 \tilde{\lambda}^4 + 1 \big) + \frac{1}{2 \pi} t_0^2 \tilde{\lambda}^2 \tilde{V} }{\big(\frac{2}{\pi} t_0^2 \tilde{\lambda}^2 \tilde{\Gamma}_0 + 1 \big) \big(\tilde{\Gamma}_0 + \frac{1}{2 \pi} t_0^2 \tilde{\lambda}^2 \big)}, \nonumber \\
V_{xp} & = & -\frac{\frac{1}{\pi} t_0^2 \tilde{\lambda}^2 \tilde{\Gamma}_0 \tilde{V} - \frac{1}{\pi} t_0^2 \tilde{\lambda}^2 \tilde{\Gamma}_0 \coth \big( \frac{1}{2 \tilde{T}} \big)}{\big(\frac{2}{\pi} t_0^2 \tilde{\lambda}^2 \tilde{\Gamma}_0 + 1 \big) \big(\tilde{\Gamma}_0 + \frac{1}{2 \pi} t_0^2 \tilde{\lambda}^2 \big)}.
\end{eqnarray}
Dividing these, we obtain expressions for the ratios of the covariance to the variances:
\begin{eqnarray}
\frac{V_{xp}}{V_x} & = & -\frac{1}{\pi}t_0^2 \tilde{\lambda}^2 \tilde{\Gamma}_0 \frac{\tilde{V} - \coth\big(\frac{1}{2\tilde{T}} \big)}{\frac{1}{2 \pi} t_0^2 \tilde{\lambda}^2 \tilde{V} \big(1 + 4 \tilde{\Gamma}_0^2 + \frac{2}{\pi} t_0^2 \tilde{\lambda}^2 \tilde{\Gamma}_0 \big) + \tilde{\Gamma}_0 \coth \big( \frac{1}{2 \tilde{T}} \big) }, \nonumber \\
\frac{V_{xp}}{V_p} & = & -\frac{1}{\pi}t_0^2 \tilde{\lambda}^2 \tilde{\Gamma}_0 \frac{\tilde{V} - \coth\big(\frac{1}{2\tilde{T}} \big)}{\tilde{\Gamma}_0 \coth \big( \frac{1}{2\tilde{T}} \big) \big(\frac{2}{\pi} t_0^2 \tilde{\lambda}^2 \tilde{\Gamma}_0 + \frac{1}{\pi^2} t_0^4 \tilde{\lambda}^4 + 1 \big) + \frac{1}{2 \pi} t_0^2 \tilde{\lambda}^2 \tilde{V} }.
\end{eqnarray}
It is easy to see that for $\tilde{\Gamma}_0 \le 0.1$ the magnitudes of these expressions are bounded from above either by the external damping $2 \tilde{\Gamma}_0$ or by the detector-induced damping $t_0^2 \tilde{\lambda}^2/\pi$, the latter being a very small quantity for the physically relevant values of the coupling and bare tunneling parameters.  The former bound is achieved when $t_0^2 \tilde{\lambda}^2 \tilde{V}/ 2 \pi \gg \tilde{\Gamma}_0 \coth (1/2 \tilde{T} )$.  Thus the ratios of the covariance to the two variances can be made as large as 20\% for very large external damping, $\tilde{\Gamma}_0 = 0.1$, low external temperature and sufficiently large bias voltage, tunneling amplitude and/or coupling parameter.  For example, when $t_0  = 0.1$, $\tilde{\lambda} = 0.1$, $\tilde{V} = 1\times 10^6$, $\tilde{T} = 0.01$ and $\tilde{\Gamma}_0 = 0.1$, one obtains $V_x = 166.5$, $V_p = 160.1$ and $V_{xp} =-31.8$, leading to covariance matrix eigenvalues of $131.3$ and $195.3$.  Note that the bias voltage has to be very large in order to satisfy the above condition for convergence to the $2 \tilde{\Gamma}_0$ bound.  In this case, the eigenvalues are large but unequal, indicating an elliptical Gaussian Wigner function.  

The above result is in stark contrast with the case of two position-coupled baths and $\eta = 0$, where one always obtains a circular state, whose variance is determined by the equipartition theorem with effective temperature $T_{\mathrm{eff}} = (\gamma_0 T + \gamma_{\mathrm{det}} T_{\mathrm{det}})/\gamma_{\mathrm{eff}}$, as discussed in the Introduction.  It is important to note that the eigenvalues are still between the equipartition theorem results for the two individual baths (in this case, $V_{\mathrm{det}} = 10^6$ for the detector bath and $V_{\mathrm{ext}} = 0.02$ for the external bath), and this is also the case for all other parameter values.  Besides, the naive picture that the detector bath temperature determines the variance in momentum and the external bath temperature sets the variance in position is clearly wrong - in fact, in this example $V_x > V_p$, even though $T_{\mathrm{det}} \gg T$.  The steady state of the oscillator can be characterized as a classical, thermomechanically squeezed state, similar to those studied in Ref.~[\onlinecite{rugar91}].  As already discussed, there are two sources for this thermomechanical noise squeezing - the interplay between the position and momentum coupled baths as well as the non-zero average back-action force due to $\eta \ne 0$.

Alternatively, if we allow $t_0$ and $\tilde{\lambda}$ to be as large as $0.5$, we can allow $\tilde{V}$ to be smaller while keeping the above convergence condition approximately valid - e.g. when $t_0 = 0.5$, $\tilde{\lambda} = 0.5$, $\tilde{V} = 10$, $\tilde{T} = 0.01$ and $\tilde{\Gamma}_0 = 0.1$, we have $V_x = 1.85$, $V_p = 1.81$, $V_{xp} =-0.162$, and eigenvalues equal to $1.67$ and $1.99$.  We have made the tunneling and coupling as large as possible, the bias voltage as small as possible (but still large compared to the oscillator frequency), the temperature as low as possible, and the damping as large as possible (but smaller than the oscillator frequency so that $Q \gg 1$).  Our intuition suggests that this is as close to a quantum regime as we can push the parameters and still maintain the validity of the Born-Markov approximation.  The variances in this case are close to the Heisenberg uncertainty limit, the covariance/variance ratios are large (about 10\%) and the eigenvalues are unequal, yet even in this extreme regime what we have is simply a very cold, classical, thermomechanically squeezed state, consistent with our expectations.

\section{\label{sec:conclusion}Conclusion}
We have studied the current noise spectrum and steady state behavior of a resonator coupled linearly to a QPC via its momentum for a wide range of system parameters.  Our spectra show clear signatures of the non-Gaussian correlations between the junction current and the back-action force on the oscillator, namely the resonance-antiresonance features at $\tilde{\omega} = \pm 1$.  These features are prominent in our case because the tunneling phase is set to a value ($\eta = -\pi/2$) where the current and back-action force are maximally correlated.  Our results are consistent with the analysis of Ref.~[\onlinecite{doiron08}], implying that, as far as the current noise is concerned, the momentum-coupled system is quite similar to the position-coupled system with the same $\eta$ due to canonical invariance, despite the presence of a position-coupled external bath, at least in the case of a weakly coupled environment ($Q \gg 1$).

Comparing our results to the experimental noise spectra obtained in Ref.~[\onlinecite{stettenheim10}], we find that inserting the experimental parameter values into our calculation fails to reproduce some important features of the experimental results, such as the high super-Poissonian values at the peaks or the sub-Poissonian noise away from them.  This breakdown of the theory at strong tunneling is expected as both the polaron transformation and the Born-Markov approximation we have used hinge on the assumption of weak tunneling.  However, if we keep the tunneling weak, and use stronger coupling and weaker damping than in the experiment, we can obtain spectra much more similar to the experimental ones, suggesting that a future theoretical approach that does not depend on the weak tunneling assumption might be much more successful in predicting the noise quantitatively.  In future work, we plan to use scattering matrix methods to treat arbitrarily strong tunneling, as proposed by Bennett et al.~\cite{bennett10}

Our study of the oscillator steady state indicates that once the detector is traced out, the oscillator obeys a classical Fokker-Planck equation, where it is coupled to two independent reservoirs.  Thus an unconditional measurement of the steady-state oscillator moments is not expected to yield any deviations from classicality.  However, the full master equation for the coupled system, Eq.~(\ref{Wigner}), clearly contains quantum terms showing that exactly half of the electronic back-action is correlated with tunneling.  Ref.~[\onlinecite{bennett08}] suggests that in order to observe these departures from the effective bath model, one needs to study the conditional evolution of the oscillator, based on a certain current measurement history, or else look at the current noise spectrum of the detector, as we have done in the present study.

Despite its classical nature, the oscillator steady state can experience significant thermomechanical noise squeezing.  At high external damping, low external temperature, and large bias voltage, tunneling amplitude and/or coupling strength, the variances in position and momentum can differ by up to about $20\%$, while still remaining between the limits set by the temperatures of the two individual baths.  The sources for this thermomechanical squeezing are the simultaneous presence of a position-coupled and a momentum-coupled bath, as well as the non-zero average back-action force on the oscillator due to $\eta = -\pi/2$.
 
To conclude, there are several future directions that one can take in order to extend the present study.  First, one could use a scattering approach to treat the case of strong tunneling and enter the parameter regime of current experiments.~\cite{bennett10}  Secondly, using scattering or some other approach, one could attempt to relax the high bias voltage ($eV/\hbar \omega_m \gg 1$) assumption and enter a regime where one might expect to see quantum signatures in the oscillator steady state as well as in the current noise spectrum.  Thirdly, one could look at the conditional evolution of the oscillator based on a certain measurement history of the QPC current, which can be highly non-thermal even in the weak tunneling limit.~\cite{bennett08}

\section*{Acknowledgements}
We thank Aashish Clerk and Alex Rimberg for helpful discussions.  This work was supported by the National Science Foundation under Grants No. DMR-0804477 and No. DMR-1104790.

\appendix

\section{\label{sec:appendixa}Polaronic Transformation}
Starting with Eq.~(\ref{orig_Ham}), we perform a polaron-like transformation on the Hamiltonian, i.e. $H \rightarrow U H U^{\dagger}$, with the unitary operator 
\begin{equation} 
U = \exp \Bigg[ - \frac{\lambda x_{\mathrm{zp}}}{2 \hbar \omega_m} \Bigg( \sum_L b_L^{\dagger} b_L - \sum_R b_R^{\dagger} b_R \Bigg) (a^{\dagger} - a) \Bigg].
\end{equation}
We make use of the Baker-Hausdorff lemma:~\cite{sakurai11}
\begin{equation}
e^{i G \xi} O e^{-i G \xi} = O + i \xi [G, O] + \frac{i^2 \xi^2}{2!} [G, [G, O]] + \cdots + \frac{i^n \xi^n}{n!} [G, [G, \dots [G, O] ]  \dots ],
\end{equation}
where $O$ is an operator, $G$ is a Hermitian operator, and $\xi$ is a real number.  In our case, we identify $\xi = \lambda x_{\mathrm{zp}} / (2 \hbar \omega_m)$ and $G = i \big( \sum_L b_L^{\dagger} b_L - \sum_R b_R^{\dagger} b_R \big) (a^{\dagger} - a)$.  Expanding to first order in the oscillator coordinates, i.e. in $\xi$, and using the canonical commutation (anti-commutation) relations for the $a^{(\dagger)}$ ($b_i^{(\dagger)}$) operators, we obtain the following results for the various terms in the original Hamiltonian: 
\begin{eqnarray}
U \Bigg( \sum_{L,R} \hbar \Omega_{LR} b_L^{\dagger} b_R \Bigg) U^{\dagger} & = & \Bigg[ 1 - \frac{\lambda x_{\mathrm{zp}}}{\hbar \omega_m} (a^{\dagger} - a) \Bigg] \sum_{L,R} \hbar \Omega_{LR} b_L^{\dagger} b_R, \\
U (\hbar \omega_m a^{\dagger} a) U^{\dagger} & = & \hbar \omega_m a^{\dagger} a + \frac{\lambda x}{2} \Bigg( \sum_L b_L^{\dagger} b_L - \sum_R b_R^{\dagger} b_R \Bigg), \\
U \Bigg[ \sum_L (\epsilon_L - \lambda x/2) b_L^{\dagger} b_L \Bigg] U^{\dagger} & = & \sum_L (\epsilon_L - \lambda x/2) b_L^{\dagger} b_L, \\
U \Bigg[ \sum_R (\epsilon_R + \lambda x/2) b_R^{\dagger} b_R \Bigg] U^{\dagger} & = & \sum_R (\epsilon_R + \lambda x/2) b_R^{\dagger} b_R.
\end{eqnarray}
To obtain the last two equations, we have dropped the second term in the Baker-Hausdorff formula as it leads to terms quartic in the $b_L$ and $b_R$ operators.  As the $L$ and $R$ reservoirs are to be combined with the $E$ and $C$ reservoirs, respectively (see comment after Eq.~(\ref{polaronlong}) below), there is negligible accumulation of electrons in these reservoirs, hence higher order terms in $b_L$ and $b_R$ do not contribute. Also, we have
\begin{equation}
U \Bigg( \sum_E \epsilon_E b_E^{\dagger} b_E + \sum_C \epsilon_C b_C^{\dagger} b_C \Bigg) U^{\dagger} = \sum_E \epsilon_E b_E^{\dagger} b_E + \sum_C \epsilon_C b_C^{\dagger} b_C,
\end{equation}
since $U$ clearly commutes with $H_{\mathrm{bath}}$.  Finally, for the remaining two terms in $H_{\mathrm{int}}$, we get
\begin{eqnarray}
U \Bigg( \sum_{E, L} \hbar \Omega_{EL} b_E^{\dagger} b_L \Bigg) U^{\dagger} & = & \Bigg[ 1 + \frac{\lambda x_{\mathrm{zp}}}{2 \hbar \omega_m} (a^{\dagger} - a) \Bigg] \sum_{E, L} \hbar \Omega_{EL} b_E^{\dagger} b_L, \\
U \Bigg( \sum_{C, R} \hbar \Omega_{CR} b_C^{\dagger} b_R \Bigg) U^{\dagger} & = & \Bigg[ 1 - \frac{\lambda x_{\mathrm{zp}}}{2 \hbar \omega_m} (a^{\dagger} - a) \Bigg] \sum_{C, R} \hbar \Omega_{CR} b_C^{\dagger} b_R,
\end{eqnarray}
where we can neglect the momentum-dependent second terms. 

Putting all the terms in $U H U^{\dagger}$ together, we see that the oscillator position-dependent terms cancel, and obtain
\begin{eqnarray}
U H U^{\dagger} & = & \hbar \omega_m a^{\dagger} a + \sum_L \epsilon_L b_L^{\dagger} b_L + \sum_R \epsilon_R b_R^{\dagger} b_R + \sum_E \epsilon_E b_E^{\dagger} b_E + \sum_C \epsilon_C b_C^{\dagger} b_C \nonumber \\
& + & \sum_{E, L} \hbar \Omega_{EL} b_E^{\dagger} b_L + \sum_{C, R} \hbar \Omega_{CR} b_C^{\dagger} b_R + \Bigg[ 1 - \frac{\lambda x_{\mathrm{zp}}}{\hbar \omega_m} (a^{\dagger} - a) \Bigg] \sum_{L,R} \hbar \Omega_{LR} b_L^{\dagger} b_R \nonumber \\
& + & \textrm{H.c. of last 3 terms}.
\label{polaronlong}
\end{eqnarray}
Finally, we assume that the lead-reservoir tunneling amplitudes $\Omega_{EL}$ and $\Omega_{CR}$ are much larger than the tunneling amplitude $\Omega_{LR}$ between the two reservoirs, so we can effectively combine the emitter and the left reservoir, and also the collector and the right reservoir.  Thus the transformed Hamiltonian can be written in the much simpler form $H = H_{\mathrm{osc}} + H_{\mathrm{bath}} + H_{\mathrm{int}}$, where
\begin{eqnarray}
H_{\mathrm{osc}} & = & \hbar \omega_m a^{\dagger} a, \nonumber \\
H_{\mathrm{bath}} & = & \sum_L \epsilon_L b_L^{\dagger} b_L + \sum_R \epsilon_R b_R^{\dagger} b_R, \nonumber \\
H_{\mathrm{int}} & = & \Bigg[ 1 - \frac{\lambda x_{\mathrm{zp}}}{\hbar \omega_m} (a^{\dagger} - a) \Bigg] \sum_{L,R} \hbar \Omega_{LR} b_L^{\dagger} b_R Y + \mathrm{H.c.}
\end{eqnarray}

\section{\label{sec:appendixb}Detailed Derivation of the Born-Markov Master Equation}
The starting point for the derivation of Eq.~(\ref{eq:master2}) is the general Born-Markov master equation
\begin{equation} 
\frac{d}{dt} \rho_{\mathrm{osc}}(t) = \frac{1}{i \hbar} [H_{\mathrm{0}}, \rho_{\mathrm{osc}}(t)] - \frac{1}{\hbar^2} \int_0^{\infty} dt' \mathrm{Tr_{bath}} \big\{
\big[H_{\mathrm{int}}, [H_{\mathrm{int}}(-t'), \rho_{\mathrm{osc}}(t) \otimes \rho_{\mathrm{bath}}] \big] \big\},
\end{equation}
where $H_{\mathrm{0}} = H_{\mathrm{osc}} + H_{\mathrm{bath}}$ and $H_{\mathrm{int}}(-t')$ is given in the interaction picture.  The derivation proceeds in exactly the same way as in Eqs.~(7.49)-(7.64) of Ref.~[\onlinecite{doiron09}], since the position or momentum dependence of the coupling is not made explicit until later.  Thus we can take Eqs.~(7.63)-(7.64) (adapted to our notation) as the starting point of our calculation:
\begin{equation}
\frac{d}{dt} \rho(\chi;t) = \frac{1}{i \hbar} [H_{\mathrm{0}}, \rho(\chi;t)] - \frac{1}{(2 \pi \hbar \Lambda)^2} \int_0^{\infty} dt' \sum_{R,L} A(\chi, R, L; t, t'),
\end{equation}
where
\begin{eqnarray}
A(\chi, R, L; t, t') & = & \big[\mathcal{T} \mathcal{T}^{\dagger}(-t') \rho(\chi;t) - \mathcal{T}^{\dagger}(-t') \rho(\chi;t) \mathcal{T} \big] e^{-i(\epsilon_L - \epsilon_R) t'/\hbar} f_R (1-f_L) \nonumber \\
& + & \big[\rho(\chi;t) \mathcal{T}(-t') \mathcal{T}^{\dagger} - \mathcal{T}^{\dagger} \rho(\chi;t) \mathcal{T}(-t') \big] e^{i(\epsilon_L - \epsilon_R)t'/\hbar} f_R(1-f_L) \nonumber \\
& + & \big[\mathcal{T}^{\dagger} \mathcal{T}(-t') \rho(\chi;t) - \mathcal{T}(-t') \rho(\chi;t) \mathcal{T}^{\dagger} \big] e^{i(\epsilon_L - \epsilon_R)t'/\hbar} f_L(1-f_R) \nonumber \\
& + & \big[\rho(\chi;t) \mathcal{T}^{\dagger}(-t') \mathcal{T} - \mathcal{T} \rho(\chi;t) \mathcal{T}^{\dagger}(-t') \big] e^{-i(\epsilon_L - \epsilon_R)t'/\hbar} f_L(1-f_R) \nonumber \\
& - & \big(e^{i \chi} - 1 \big) \big[\mathcal{T} \rho(\chi;t) \mathcal{T}^{\dagger}(-t') e^{-i(\epsilon_L - \epsilon_R)t'/\hbar} \big] f_L(1-f_R) \nonumber \\
& - & \big(e^{i \chi} - 1 \big) \big[\mathcal{T}(-t') \rho(\chi;t) \mathcal{T}^{\dagger} e^{i(\epsilon_L - \epsilon_R)t'/\hbar} \big] f_L(1-f_R) \nonumber \\
& - & \big(e^{-i \chi} - 1 \big) \big[\mathcal{T}^{\dagger} \rho(\chi;t) \mathcal{T}(-t') e^{i(\epsilon_L - \epsilon_R)t'/\hbar} \big] f_R(1-f_L) \nonumber \\
& - & \big(e^{-i \chi} - 1 \big) \big[\mathcal{T}^{\dagger}(-t') \rho(\chi;t) \mathcal{T} e^{-i(\epsilon_L - \epsilon_R)t'/\hbar} \big] f_R(1-f_L).
\label{eq:A}
\end{eqnarray}
In this equation, $\epsilon_L$($\epsilon_R$) and $f_L$($f_R$) are the energy levels and Fermi finctions of the left (right) reservoir and $\mathcal{T}$ is defined as $\mathcal{T} = t_0 + t_1 e^{i \eta} \hat{p}$ such that
\begin{equation}
\hat{T} = \frac{1}{2 \pi \Lambda} \mathcal{T} Y^{\dagger}.
\end{equation}
Using the interaction picture expression for the harmonic oscillator momentum,
\begin{equation}
\hat{p}(t) = -m \omega_m \hat{x} \sin(\omega_m t) + \hat{p} \cos(\omega_m t),
\end{equation}
we can write
\begin{equation}
\mathcal{T}(t) = t_0 + \frac{t_1}{2} e^{i \eta} (\hat{p} + i m \omega_m \hat{x}) e^{i \omega_m t} + \frac{t_1}{2} e^{i \eta} (\hat{p} - i m \omega_m \hat{x}) e^{-i \omega_m t},
\end{equation} 
with an analogous expression for $\mathcal{T}^{\dagger}(t)$.  Substituting these into Eq.~(\ref{eq:A}), we first perform the integration over $t'$ using
\begin{equation}
\int_0^{\infty} dt e^{\pm i \omega t} = \pi \delta(\omega) \pm i \mathrm{pv} \bigg( \frac{1}{\omega} \bigg).
\label{Fourierint}
\end{equation}
In the case of position-dependent coupling, the principal value term leads to a term that renormalizes the oscillator frequency plus another negligible term in the Caldeira-Leggett equation.  In our case, we expect it to lead to oscillator mass renormalization (i.e. an extra term $\propto \hat{p}^2$) plus a similar negligible term, so we drop the principal value part in complete analogy with the position coupling case.  Next, we replace the summations over discrete energy levels by integrals over energy with constant density of states $\Lambda$:
\begin{equation}
\sum_i \dots \rightarrow \int_0^{\infty} d \epsilon_i \Lambda \dots,
\end{equation}
where $i = L,R$.  The result of the integration over the reservoir Fermi functions can be expressed in terms of the tunneling rates $\Gamma_{\pm}(E)$:
\begin{eqnarray}
h \Gamma_{+}(E) & = & \int_0^\infty d \epsilon |t_0|^2 f(\epsilon - \mu_L) [1 - f(\epsilon - \mu_R + E)],  \\
h \Gamma_{-}(E) & = & \int_0^\infty d \epsilon |t_0|^2 f(\epsilon - \mu_R) [1 - f(\epsilon - \mu_L + E)], 
\end{eqnarray}
where $\mu_i$ is the chemical potential in reservoir $i$.  After a somewhat lengthy but straightforward collection of terms, one arrives at Eq.~(\ref{eq:master2}).

\section{\label{sec:appendixc}Detailed Solution of the Master Equation to Find the Current Noise Spectrum}
In this appendix, we solve for the time dependence of the cumulants $\langle \langle x N \rangle \rangle$ and $\langle \langle p^2 N \rangle \rangle$, and integrate the MacDonald formula to obtain exact analytical expressions for the non-Poissonian current noise spectrum $\Delta \bar{S}_I(\omega)$ in Eq.~(\ref{eq:deltaSI}).  First we solve for $\langle \langle x N \rangle \rangle$.  Using the method of taking derivatives with respect to $\chi$ and then tracing the master equation over the oscillator degrees of freedom, we obtain two coupled first-order differential equations for $\langle \langle x N \rangle \rangle$ and $\langle \langle p N \rangle \rangle$:
\begin{eqnarray}
\frac{d}{dt} \langle \langle x N \rangle \rangle & = & -2 m^2 \omega_m^2 \tilde{\gamma}_{+} \langle \langle x N \rangle \rangle + \frac{1}{m} \langle \langle p N \rangle \rangle + \frac{\hbar e V t_0 t_1}{h} - m^2 \omega_m^2 \tilde{\gamma}_{+} \langle x \rangle \nonumber \\
& - & \frac{i \hbar e V t_1^2}{h} \langle p \rangle - \frac{2 m^2 \omega_m^2 \tilde{\gamma}_{+}}{\hbar} \frac{t_0}{t_1} \big(\langle x^2 \rangle - \langle x \rangle^2 \big) + \frac{e V t_1^2}{h} (\langle x p^2 \rangle - \langle x \rangle \langle p^2 \rangle \big), \nonumber \\
\frac{d}{dt} \langle \langle p N \rangle \rangle & = & -m \omega_m^2 \langle \langle x N \rangle \rangle - 2 \tilde{\gamma}_0 \langle \langle p N \rangle \rangle + i m^2 \omega_m^2 \tilde{\gamma}_{+} \frac{t_0}{t_1} - m^2 \omega_m^2 \tilde{\gamma}_{+} \langle p \rangle \nonumber \\
& - & \frac{2 m^2 \omega_m^2 \tilde{\gamma}_{+}}{\hbar} \frac{t_0}{t_1} \big(\langle xp \rangle - \langle x \rangle \langle p \rangle \big) + \frac{e V t_1^2}{h} \big( \langle p^3 \rangle - \langle p \rangle \langle p^2 \rangle \big).
\label{eq:fom}
\end{eqnarray}
We replace all averages that do not contain $N$ with their stationary values in order that the current-current correlation function $\langle I(t + t') I(t) \rangle$ be independent of $t$.  Eliminating $\langle \langle p N \rangle \rangle$ and imposing the boundary conditions $\langle \langle x N(t=0) \rangle \rangle = \langle \langle p N(t=0) \rangle \rangle = 0$, we obtain the following initial value problem for $y = \langle \langle x N \rangle \rangle$:
\begin{eqnarray}
\ddot{y} + a \dot{y} + b y & = & c, \nonumber \\
y(0) & = & 0, \nonumber \\ 
\dot{y}(0) & = & d,
\end{eqnarray}
where 
\begin{eqnarray}
a & = & 2 \big( m^2 \omega_m^2 \tilde{\gamma}_{+} + \tilde{\gamma}_0 \big), \nonumber \\
b & = & \omega_m^2 \big(1 + 4 m^2 \tilde{\gamma}_{+} \tilde{\gamma}_0 \big), \nonumber \\
c & = & \frac{2 \hbar eV t_0 t_1 \tilde{\gamma}_0}{h} + i m \omega_m^2 \tilde{\gamma}_{+} \frac{t_0}{t_1} - 2 m^2 \omega_m^2 \tilde{\gamma}_{+} \tilde{\gamma}_0 \langle x \rangle - m \omega_m^2 \tilde{\gamma}_{+} \langle p \rangle \nonumber \\
& - & \frac{2 i \hbar eV t_1^2 \tilde{\gamma}_0}{h} \langle p \rangle - \frac{4 m^2 \omega_m^2 \tilde{\gamma}_{+} \tilde{\gamma}_0}{\hbar} \frac{t_0}{t_1} \big( \langle x^2 \rangle - \langle x \rangle^2 \big) - \frac{2 m \omega_m^2 \tilde{\gamma}_{+}}{\hbar} \frac{t_0}{t_1} \big( \langle xp \rangle - \langle x \rangle \langle p \rangle \big) \nonumber \\
& + & \frac{2 e V t_1^2 \tilde{\gamma}_0}{h} \big( \langle xp^2 \rangle - \langle x \rangle \langle p^2 \rangle \big) + \frac{e V t_1^2}{m h} \big( \langle p^3 \rangle - \langle p \rangle \langle p^2 \rangle \big), \nonumber \\
d & = & \frac{\hbar e V t_0 t_1}{h} - m^2 \omega_m^2 \tilde{\gamma}_{+} \langle x \rangle - \frac{i \hbar e V t_1^2}{h} \langle p \rangle \nonumber \\
& - & \frac{2 m^2 \omega_m^2 \tilde{\gamma}_{+}}{\hbar} \frac{t_0}{t_1} \big( \langle x^2 \rangle - \langle x \rangle^2 \big) + \frac{e V t_1^2}{h} \big( \langle x p^2 \rangle - \langle x \rangle \langle p^2 \rangle \big).
\end{eqnarray}
This is a simple second-order linear inhomogeneous differential equation with constant coefficients, whose general solution is 
\begin{equation}
y(t) = A e^{r_1 t} + B e^{r_2 t} + \frac{c}{b},
\label{eq:xm}
\end{equation}
where 
\begin{equation}
r_{1,2} = \frac{-a \pm \sqrt{a^2 - 4b}}{2}
\end{equation}
are the roots of the auxiliary equation (assumed to be distinct). Applying the initial conditions, we find
\begin{eqnarray}
A & = & \frac{r_2 c + b d}{b(r_1 - r_2)}, \nonumber \\
B & = & - \frac{r_1 c + b d}{b (r_1 - r_2)}.
\end{eqnarray}
We substitute this solution into Eq.~(\ref{eq:deltaSI}) to obtain the first term in $\Delta \bar{S}_I$.  The resulting integrals converge provided that $\mathrm{Re}(r_{1,2}) < 0$, which is easily seen to be the case.  The term 
\begin{equation}
\int_0^{\infty} dt \sin \omega t \bigg( \frac{c}{b} \bigg)
\end{equation}
is not integrable, but can be evaluated by the method of Cesaro summation:
\begin{equation}
\int_0^\infty K \sin \omega t dt = \frac{K}{\omega}.
\end{equation}
We trust this method as it removes the discontinuity in the noise spectrum at $\omega = 0$ and also makes the noise go to zero at large frequencies instead of having a zero level that's very large in magnitude and varies randomly with the choice of parameters.  Finally, one obtains for the first term in Eq.~(\ref{eq:deltaSI}), i.e.~for the contribution from $\langle \langle x N \rangle \rangle$:
\begin{equation}
\Delta \bar{S}_I^{\textrm{(1st term)}} = - \frac{8 m^2 \omega_m^2 \tilde{\gamma}_{+}}{\hbar} \frac{t_0}{t_1} e^2 \omega \bigg[ \frac{A \omega}{r_1^2 + \omega^2} + \frac{B \omega}{r_2^2 + \omega^2} + \frac{c}{b} \bigg( \frac{1}{\omega} \bigg) \bigg].
\label{1storder}
\end{equation}

In a similar way we can solve for $\langle \langle p^2 N \rangle \rangle$, needed to calculate the second term in the non-Poissonian current noise.  We first obtain three coupled linear differential equations for the cumulants that are second-order in the oscillator variables:
\begin{eqnarray}
\frac{d}{dt} \langle \langle x^2 N \rangle \rangle & = & c_1 \langle \langle x^2 N \rangle \rangle + c_2 \langle \langle (xp) N \rangle \rangle + c_3 \langle \langle x N \rangle \rangle + c_4, \nonumber \\
\frac{d}{dt} \langle \langle (xp) N \rangle \rangle & = & c_5 \langle \langle x^2 N \rangle \rangle + c_6 \langle \langle (xp) N \rangle \rangle + c_7 \langle \langle p^2 N \rangle \rangle + c_8 \langle \langle p N \rangle \rangle + c_9, \nonumber \\
\frac{d}{dt} \langle \langle p^2 N \rangle \rangle & = & c_{10} \langle \langle (xp) N \rangle \rangle + c_{11} \langle \langle p^2 N \rangle \rangle + c_{12},
\end{eqnarray}
where we have defined the constants $c_i$ as
\begin{eqnarray}
c_1 & = & -4 m^2 \omega_m^2 \tilde{\gamma}_+, \nonumber \\
c_2 & = & \frac{2}{m}, \nonumber \\
c_3 & = & \frac{2 \hbar t_0 t_1 e V}{h}, \nonumber \\
c_4 & = & \frac{2 \hbar t_0 t_1 e V}{h} \langle x \rangle - 2 m^2 \omega_m^2 \tilde{\gamma}_+ \langle x^2 \rangle - \frac{2 i \hbar t_1^2 eV}{h} \langle xp \rangle \nonumber \\
& + & \frac{2 m^2 \omega_m^2 \tilde{\gamma}_+}{\hbar} \frac{t_0}{t_1} \big( \langle x \rangle \langle x^2 \rangle - \langle x^3 \rangle \big) + \frac{t_1^2 eV}{h} \big( \langle x^2 p^2 \rangle - \langle x^2 \rangle \langle p^2 \rangle \big), \nonumber \\
c_5 & = & -m \omega_m^2, \nonumber \\
c_6 & = & -2 \big( m^2 \omega_m^2 \tilde{\gamma}_+ + \tilde{\gamma}_0 \big), \nonumber \\
c_7 & = & \frac{1}{m}, \nonumber \\
c_8 & = & \frac{\hbar t_0 t_1 eV}{h}, \nonumber \\
c_9 & = & i \hbar m^2 \omega_m^2 \tilde{\gamma}_+ + im^2 \omega_m^2 \tilde{\gamma}_+ \frac{t_0}{t_1} \langle x \rangle + \frac{\hbar t_0 t_1 eV}{h} \langle p \rangle - 2 m^2 \omega_m^2 \tilde{\gamma}_+ \langle xp \rangle \nonumber \\
& - & \frac{i \hbar t_1^2 eV}{h} \langle p^2 \rangle - \frac{2 m^2 \omega_m^2 \tilde{\gamma}_+}{\hbar} \frac{t_0}{t_1} \big( \langle x^2 p \rangle - \langle x \rangle \langle xp \rangle \big) + \frac{t_1^2 eV}{h} \big( \langle xp^3 \rangle - \langle xp \rangle \langle p^2 \rangle \big), \nonumber \\
c_{10} & = & -2 m \omega_m^2, \nonumber \\
c_{11} & = & -4 \tilde{\gamma}_0, \nonumber \\
c_{12} & = & 2 i m^2 \omega_m^2 \tilde{\gamma}_+ \frac{t_0}{t_1} \langle p \rangle - 2 m^2 \omega_m^2 \tilde{\gamma}_+ \langle p^2 \rangle - \frac{2 m^2 \omega_m^2 \tilde{\gamma}_+}{\hbar} \frac{t_0}{t_1} \big( \langle xp^2 \rangle - \langle x \rangle \langle p^2 \rangle \big) \nonumber \\
& + & \frac{t_1^2 eV}{h} \big( \langle p^4 \rangle - {\langle p^2 \rangle}^2 \big).
\end{eqnarray}
Eliminating $\langle \langle x^2 N \rangle \rangle$ and $\langle \langle (xp) N \rangle \rangle$ and substituting for $\langle \langle xN \rangle \rangle$ and $\langle \langle pN \rangle \rangle$ using Eqs.~(\ref{eq:xm}) and (\ref{eq:fom}), we get a third-order linear inhomogeneous ordinary differential equation for $z = \langle \langle p^2 N \rangle \rangle$:
\begin{equation} 
\frac{d^3 z}{dt^3} + \alpha \frac{d^2 z}{dt^2} + \beta \frac{dz}{dt} + \gamma z = \mu e^{r_1 t} + \nu e^{r_2 t} + \rho,
\end{equation}
where we have defined the constants
\begin{eqnarray}
\alpha & = & -(c_1 + c_6 +c_{11}), \nonumber \\
\beta & = & c_1 c_6 + c_1 c_{11} - c_2 c_5 + c_6 c_{11} - c_7 c_{10}, \nonumber \\
\gamma & = & - c_1 c_6 c_{11} + c_1 c_7 c_{10} + c_2 c_5 c_{11}, \nonumber \\
\mu & = & A \big( c_3 c_5 c_{10} + c_8 c_{10} m r_1^2 + 2 c_8 c_{10} m^3 \omega_m^2 \tilde{\gamma}_+ r_1 \nonumber \\
& - & c_1 c_8 c_{10} m r_1 - 2 c_1 c_8 c_{10} m^3 \omega_m^2 \tilde{\gamma}_+ \big), \nonumber \\
\nu & = & B \big( c_3 c_5 c_{10} + c_8 c_{10} m r_2^2 + 2 c_8 c_{10} m^3 \omega_m^2 \tilde{\gamma}_+ r_2 \nonumber \\
& - & c_1 c_8 c_{10} m r_2 - 2 c_1 c_8 c_{10} m^3 \omega_m^2 \tilde{\gamma}_+ \big), \nonumber \\
\rho & = & c_3 c_5 c_{10} \frac{c}{b} - 2 c_1 c_8 c_{10} m^3 \omega_m^2 \tilde{\gamma}_+ \frac{c}{b} + c_1 c_6 c_{12} + c_1 c_8 c_{10} m d \nonumber \\
& - & c_1 c_9 c_{10} - c_2 c_5 c_{12} + c_4 c_5 c_{10},
\end{eqnarray}
and $b$, $c$, $r_1$, $r_2$, $A$ and $B$ are as defined above.  The general solution is easily found to be
\begin{equation}
z(t) = C e^{\rho_1 t} + D e^{\rho_2 t} + E e^{\rho_3 t} + M e^{r_1 t} + N e^{r_2 t} + \frac{\rho}{\gamma},
\label{eq:p2N}
\end{equation}
where $\rho_i$ are the three roots of the auxiliary equation $\rho^3 + \alpha \rho^2 + \beta \rho + \gamma = 0$ (assumed to be distinct), the constants $C$, $D$ and $E$ are to be found by imposing initial conditions, and
\begin{eqnarray}
M & = & \frac{\mu}{r_1^3 + \alpha r_1^2 + \beta r_1 + \gamma}, \nonumber \\
N & = & \frac{\nu}{r_2^3 + \alpha r_2^2 + \beta r_2 + \gamma}.
\end{eqnarray}
Assuming $\alpha^2 - 3 \beta \neq 0$ and using the cubic equation formula, we find
\begin{eqnarray}
\rho_1 & = & - \frac{\alpha}{3} - \frac{R}{3} -\frac{\alpha^2 - 3 \beta}{3 R}, \nonumber \\
\rho_2 & = & - \frac{\alpha}{3} + \frac{R(1 + i \sqrt{3})}{6} + \frac{(1 - i \sqrt{3})(\alpha^2 - 3 \beta)}{6 R}, \nonumber \\
\rho_3 & = & - \frac{\alpha}{3} + \frac{R(1 - i \sqrt{3})}{6} + \frac{(1 + i \sqrt{3})(\alpha^2 - 3 \beta)}{6 R},
\end{eqnarray}
where 
\begin{eqnarray}
Q & = & \sqrt{(2 \alpha^3 - 9 \alpha \beta + 27 \gamma)^2 - 4(\alpha^2 - 3 \beta)^3}, \nonumber \\
R & = & \sqrt[3]{\frac{1}{2} \bigg(Q + 2 \alpha^3 - 9 \alpha \beta + 27 \gamma \bigg)},
\end{eqnarray}
and any determination of the complex-valued square and cube roots can be used.  From the boundary conditions $\langle \langle x^2 N \rangle \rangle = \langle \langle (xp) N \rangle \rangle = \langle \langle p^2 N \rangle \rangle = \langle \langle x N \rangle \rangle = \langle \langle p N \rangle \rangle = 0$ at $t = 0$ we obtain the appropriate initial conditions:
\begin{eqnarray}
z(0) & = & 0, \nonumber \\
\dot{z}(0) & = & \delta, \nonumber \\
\ddot{z}(0) & = & \epsilon,
\end{eqnarray}
where
\begin{eqnarray}
\delta & = & c_{12}, \nonumber \\
\epsilon & = & c_9 c_{10} + c_{11} c_{12}.
\end{eqnarray}
Applying these initial conditions and solving the resulting system of equations, we obtain
\begin{eqnarray}
C & = & \frac{M(\rho_2 \rho_3 - r_1 \rho_2 - r_1 \rho_3 + r_1^2) + N(\rho_2 \rho_3 - r_2 \rho_2 - r_2 \rho_3 + r_2^2) + \rho \rho_2 \rho_3/\gamma + \delta \rho_2 + \delta \rho_3 - \epsilon}{(\rho_1 - \rho_2)(\rho_3 - \rho_1)}, \nonumber \\
D & = & \frac{M(\rho_1 \rho_3 - r_1 \rho_1 - r_1 \rho_3 + r_1^2) + N(\rho_1 \rho_3 - r_2 \rho_1 - r_2 \rho_3 + r_2^2) + \rho \rho_1 \rho_3/\gamma + \delta \rho_1 + \delta \rho_3 - \epsilon}{(\rho_1 - \rho_2)(\rho_2 - \rho_3)}, \nonumber \\
E & = & \frac{M(\rho_1 \rho_2 - r_1 \rho_1 - r_1 \rho_2 + r_1^2) + N(\rho_1 \rho_2 - r_2 \rho_1 - r_2 \rho_2 + r_2^2) + \rho \rho_1 \rho_2/\gamma + \delta \rho_1 + \delta \rho_2 - \epsilon}{(\rho_2 - \rho_3)(\rho_3 - \rho_1)}. \nonumber \\
\end{eqnarray}
Having determined all the relevant constants, we finally substitute Eq.~(\ref{eq:p2N}) into the MacDonald formula, Eq.~(\ref{eq:deltaSI}), to obtain the second term in $\Delta \bar{S}_I$.  The integration proceeds in exactly the same way as in the calculation of the first term.  One has to check that the roots $\rho_1$, $\rho_2$ and $\rho_3$ have negative real parts, which is best done numerically as the analytical expressions are rather complicated.  Finally, we obtain for the second term in Eq.~(\ref{eq:deltaSI}), i.e.~the contribution from $\langle \langle p^2 N \rangle \rangle$:
\begin{equation}
\Delta \bar{S}_I^{\textrm{(2nd term)}} = \frac{4 e^3 V t_1^2}{h} \omega \bigg[ \frac{C \omega}{\rho_1^2 + \omega^2} + \frac{D \omega}{\rho_2^2 + \omega^2} + \frac{E \omega}{\rho_3^2 + \omega^2} + \frac{M \omega}{r_1^2 + \omega^2} + \frac{N \omega}{r_2^2 + \omega^2} + \frac{\rho}{\gamma} \bigg(\frac{1}{\omega} \bigg) \bigg].
\label{2ndorder}
\end{equation}

In the above calculations, we needed to know the $N$-independent oscillator moments, $\langle x^i p^j \rangle$, up to fourth order.  These can be found easily by solving systems of linear equations successively, starting with the first-order moments, then going to second order, etc.  The equations for the moments are obtained by miltiplying Eq.~(\ref{eq:master2}) by $x^i p^j$, setting $\chi = 0$, tracing over the oscillator degrees of freedom, and finally setting $d \langle x^i p^j \rangle /dt = 0$ since all oscillator moments are stationary as already discussed.  The moment equations are given below.
\\
First-order moments:
\begin{eqnarray}
\frac{d}{dt} \langle x \rangle & = & \frac{\hbar t_0 t_1 eV}{h} -2 m^2 \omega_m^2 \tilde{\gamma}_+ \langle x \rangle + \frac{1}{m} \langle p \rangle = 0 \nonumber \\
\frac{d}{dt} \langle p \rangle & = & -m \omega_m^2 \langle x \rangle - 2 \tilde{\gamma}_0 \langle p \rangle = 0
\label{firstordermoments}
\end{eqnarray}
\\
Second-order moments:
\begin{eqnarray}
\frac{d}{dt} \langle x^2 \rangle & = & -\frac{i \hbar}{m} + \frac{\hbar^2 t_1^2 eV}{h} + \frac{2 \hbar t_0 t_1 eV}{h} \langle x \rangle - 4 m^2 \omega_m^2 \tilde{\gamma}_+ \langle x^2 \rangle + \frac{2}{m} \langle xp \rangle = 0 \nonumber \\  
\frac{d}{dt} \langle xp \rangle & = & i \hbar \big( m^2 \omega_m^2 \tilde{\gamma}_+ + \tilde{\gamma}_0 \big) + \frac{\hbar t_0 t_1 eV}{h} \langle p \rangle - m \omega_m^2 \langle x^2 \rangle -2 \big(m^2 \omega_m^2 \tilde{\gamma}_+ + \tilde{\gamma}_0 \big) \langle xp \rangle + \frac{1}{m} \langle p^2 \rangle = 0 \nonumber \\
\frac{d}{dt} \langle p^2 \rangle & = & 2 D_0 + i \hbar m \omega_m^2 -2 m \omega_m^2 \langle xp \rangle - 4 \tilde{\gamma}_0 \langle p^2 \rangle = 0
\label{secondordermoments}
\end{eqnarray}
\\
Third-order moments:
\begin{eqnarray}
\frac{d}{dt} \langle x^3 \rangle & = & 3 \bigg( \frac{\hbar^2 t_1^2 eV}{h} - \frac{i \hbar}{m} \bigg) \langle x \rangle + \frac{3 \hbar t_0 t_1 eV}{h} \langle x^2 \rangle - 6 m^2 \omega_m^2 \tilde{\gamma}_+ \langle x^3 \rangle + \frac{3}{m} \langle x^2 p \rangle = 0 \nonumber \\
\frac{d}{dt} \langle x^2 p \rangle & = & 2 i \hbar \big(m^2 \omega_m^2 \tilde{\gamma}_+ + \tilde{\gamma}_0 \big) \langle x \rangle + \bigg(\frac{\hbar^2 t_1^2 eV}{h} - \frac{i \hbar}{m} \bigg) \langle p \rangle + \frac{2 \hbar t_0 t_1 eV}{h} \langle xp \rangle \nonumber \\
& - & m \omega_m^2 \langle x^3 \rangle -2 \big(2 m^2 \omega_m^2 \tilde{\gamma}_+ + \tilde{\gamma}_0 \big) \langle x^2 p \rangle + \frac{2}{m} \langle xp^2 \rangle = 0 \nonumber \\
\frac{d}{dt} \langle xp^2 \rangle & = & \big( i \hbar m \omega_m^2 + 2 D_0 \big) \langle x \rangle + 2 i \hbar \big( m^2 \omega_m^2 \tilde{\gamma}_+ + \tilde{\gamma}_0 \big) \langle p \rangle + \frac{\hbar t_0 t_1 eV}{h} \langle p^2 \rangle \nonumber \\
& - & 2 m \omega_m^2 \langle x^2 p \rangle - 2 \big( m^2 \omega_m^2 \tilde{\gamma}_+ + 2 \tilde{\gamma}_0 \big) \langle xp^2 \rangle + \frac{1}{m} \langle p^3 \rangle = 0 \nonumber \\
\frac{d}{dt} \langle p^3 \rangle & = & 3 \big( i \hbar m \omega_m^2 + 2 D_0 \big) \langle p \rangle - 3 m \omega_m^2 \langle xp^2 \rangle - 6 \tilde{\gamma}_0 \langle p^3 \rangle = 0
\end{eqnarray}
\\
Fourth-order moments:
\begin{eqnarray}
\frac{d}{dt} \langle x^4 \rangle & = & 6 \bigg( \frac{\hbar^2 t_1^2 eV}{h} - \frac{i \hbar}{m} \bigg) \langle x^2 \rangle + \frac{4 \hbar t_0 t_1 eV}{h} \langle x^3 \rangle - 8 m^2 \omega_m^2 \tilde{\gamma}_+ \langle x^4 \rangle + \frac{4}{m} \langle x^3 p \rangle = 0 \nonumber \\
\frac{d}{dt} \langle x^3 p \rangle & = & 3 i \hbar \big( m^2 \omega_m^2 \tilde{\gamma}_+ + \tilde{\gamma}_0 \big) \langle x^2 \rangle + 3 \bigg( \frac{\hbar^2 t_1^2 eV}{h} - \frac{i \hbar}{m} \bigg) \langle xp \rangle + \frac{3 \hbar t_0 t_1 eV}{h} \langle x^2 p \rangle \nonumber \\
& - & m \omega_m^2 \langle x^4 \rangle - 2 \big( 3 m^2 \omega_m^2 \tilde{\gamma}_+ + \tilde{\gamma}_0 \big) \langle x^3 p \rangle + \frac{3}{m} \langle x^2 p^2 \rangle = 0 \nonumber \\
\frac{d}{dt} \langle x^2 p^2 \rangle & = & \big( i \hbar m \omega_m^2 + 2 D_0 \big) \langle x^2 \rangle + 4 i \hbar \big(m^2 \omega_m^2 \tilde{\gamma}_+ + \tilde{\gamma}_0 \big) \langle xp \rangle + \bigg( \frac{\hbar^2 t_1^2 eV}{h} - \frac{i \hbar}{m} \bigg) \langle p^2 \rangle \nonumber \\
& + & \frac{2 \hbar t_0 t_1 eV}{h} \langle xp^2 \rangle - 2 m \omega_m^2 \langle x^3 p \rangle - 4 \big(m^2 \omega_m^2 \tilde{\gamma}_+ + \tilde{\gamma}_0 \big) \langle x^2 p^2 \rangle + \frac{2}{m} \langle xp^3 \rangle = 0 \nonumber \\
\frac{d}{dt} \langle xp^3 \rangle & = & 3 \big( i \hbar m \omega_m^2 + 2 D_0 \big) \langle xp \rangle + 3 i \hbar \big(m^2 \omega_m^2 \tilde{\gamma}_+ + \tilde{\gamma}_0 \big) \langle p^2 \rangle + \frac{\hbar t_0 t_1 eV}{h} \langle p^3 \rangle \nonumber \\
& - & 3 m \omega_m^2 \langle x^2 p^2 \rangle - 2 \big(m^2 \omega_m^2 \tilde{\gamma}_+ + 3 \tilde{\gamma}_0 \big) \langle xp^3 \rangle + \frac{1}{m} \langle p^4 \rangle = 0 \nonumber \\
\frac{d}{dt} \langle p^4 \rangle & = & 6 \big(i \hbar m \omega_m^2 + 2 D_0 \big) \langle p^2 \rangle - 4 m \omega_m^2 \langle xp^3 \rangle - 8 \tilde{\gamma}_0 \langle p^4 \rangle = 0 
\end{eqnarray}
\\
Importantly, the equations for the moments of a given order only involve moments of lower order, which have already been calculated and become part of the constant vector in the resulting matrix equation.  Thus there is no need for truncation or use of a semi-classical approximation in our case.

\end{document}